\documentstyle[12pt]{article}
\input epsf

\begin{document}
\baselineskip=15pt

\newcommand{\bc}{\begin{center}}
\newcommand{\ec}{\end{center}}
\newcommand{\be}{\begin{equation}}
\newcommand{\ee}{\end{equation}}
\newcommand{\PP}{\mbox{$\Psi$}}
\newcommand{\wdg}{\mbox{$\wedge$}}
\newcommand{\bq}{\begin{eqnarray}}
\newcommand{\eq}{\end{eqnarray}}
\newcommand{\del}{\mbox{$\delta$}}
\newcommand{\AC}{\mbox{${\bf {\cal A}}$}}
\newcommand{\BA}{\mbox{${\bf A}$}}
\newcommand{\BB}{\mbox{${\bf B}$}}
\newcommand{\EE}{\mbox{${\bf E}$}}
\newcommand{\A}{\mbox{${\bf A}$}}
\newcommand{\ba}{\mbox{${\bf a}$}}

\newcommand{\DC}{\mbox{${\bf {\cal D}}$}}
\newcommand{\xx}{{\bf x}}
\newcommand{\yy}{{\bf y}}
\newcommand{\Lam}{\mbox{${\Lambda}$}}
\newcommand{\lam}{\mbox{${\lambda}$}}
\newcommand{\eps}{\mbox{$\epsilon$}}
\newcommand{\vol}{\mbox{$d^{3}{\bf x}$}}
\newcommand{\Sc}{Schr\"odinger\ }
\newcommand{\p}{\mbox{$\varphi$}}
\newcommand{\der}{\partial}
\newcommand{\pp}{{\bf p}}
\newcommand{\qq}{{\bf q}}
\newcommand{\nn}{\nonumber\\}

\begin{titlepage}
\rightline{DTP 98-45}
\vskip1in
\begin{center}
{\Large{\bf Yang-Mills beta-function from a large-distance expansion
of the \Sc functional}}
\end{center}
\vskip1in
\begin{center}
{\large {\bf
Paul Mansfield}$^\dagger$ and
{\bf Marcos Sampaio}$^{\dagger \,\, \ddagger}$}\\
\vskip0.5cm
$\dagger$ Department of Mathematical Sciences\\
University of Durham\\
South Road\\
Durham, DH1 3LE, England 
\end{center}

\begin{center}
$\ddagger$ Universidade Federal de Minas Gerais\\
ICEx - Physics Department\\
P.O.Box 702, 30161-970, Belo Horizonte - MG\\
Brazil 
\end{center}
\vskip0.5cm
\begin{center}
{\it P.R.W.Mansfield@durham.ac.uk, msampaio@fisica.ufmg.br}
\end{center}
\vskip1in
\begin{abstract}
\noindent
For slowly varying fields the Yang-Mills \Sc functional
can be expanded in terms of local functionals. We
show how analyticity in a complex scale parameter enables the \Sc functional for arbitrarily varying fields to be reconstructed from this expansion. We also construct the form of the \Sc equation that determines the coefficients.  Solving this in powers of the coupling reproduces the 
results of the `standard' perturbative solution of the functional
\Sc equation which we also describe. In particular the usual result for the 
beta-function is obtained illustrating how 
analyticity enables the effects of rapidly varying fields
to be computed from the behaviour of slowly varying ones.
\end{abstract}

\end{titlepage}

\section{Introduction}

Whilst elementary discussions of quantum field theory often develop
the subject along the same lines as quantum mechanics   
by postulating equal time commutation relations for operators that act on 
a state space and a Hamiltonian that generates infinitesimal time-translations,
this formalism is in practice soon abandoned in favour of the
functional integrals that formally solve for the time-dependence.
Nonetheless the parallel between field theory and \Sc's formulation of
wave mechanics has been developed further, principally by 
Symanzik, who showed that `in every renormalisable theory, the
\Sc representation exists' for which the field operator is diagonal 
and that `in this representation, a \Sc 
equation holds', \cite{SYMANZIK}-\cite{REV5}. This representation provides  
a framework for novel approaches to the solution of quantum field 
theories \cite{Hat}-\cite{Maeda}, for example in the use of variational principles to 
model states and describe non-perturbative phenomena. It is also implicit
in much recent work on field theories defined on space-times with boundaries,
see for example \cite{witten}.

Although Symanzik's work was limited to perturbation theory 
we may assume that the \Sc equation has an existence that transcends this,
since without an equation determining the time-development of states,
physics would have no predictive power. In \cite{Paul1} it was
shown that the \Sc equation itself can be reduced to coupled {\it algebraic}
equations that can be solved without recourse to perturbation theory, \cite{forth}. Consider the vacuum functional. This is the exponential of
a functional of the eigenvalue of the field operator on the 
quantisation surface. In perturbation theory this a
sum of connected Feynman diagrams. This functional  
is non-local, but 
it will reduce to the integral of an infinite sum of local terms
when it is evaluated for fields that vary slowly on the scale of 
the mass of the lightest particle in the theory.
Each term will depend on a finite power of the field and its derivatives at a 
single point, $\bf x$, on the quantisation surface. The use of local expansions to describe Yang-Mills wave-functionals was introduced by Greensite,
\cite{Green}.
Although this expansion is appropriate  for slowly varying fields the vacuum functional for arbitrarily varying fields can be reconstructed from it because the functional obtained
by a scale-transformation in which ${\bf x}\rightarrow {\bf x}/\sqrt \rho$
is analytic in the complex $\rho$-plane with the negative real axis removed,
\cite{PM}.
Thus Cauchy's theorem relates rapidly varying fields (small $\rho$)
to slowly varying ones (large $\rho$). Similar considerations enable
 the \Sc equation,
which because it has an ultra-violet cut off involves rapidly varying fields, to be turned into an equation acting directly on the derivative expansion,
determining the coefficients of that expansion, \cite{Paul1}. It is essential to show 
that this approach does not conflict with what we already know from
perturbation theory where that is applicable. In \cite{PMJ} we showed that
for $\phi^4$ theory in 1+1 dimensions the short-distance behaviour displayed
in the leading order counter-terms was correctly reproduced by our
formulation of the \Sc equation when that was solved by perturbation theory
in the coupling, even though our approach is based on a large-distance expansion. In this paper we show that the same is true for Yang-Mills theory
in 3+1 dimensions. In particular we obtain the usual result for the beta-function by using a local expansion of the \Sc functional.
The beta-function is a result of rapidly varying
field configurations, nonetheless analyticity enables us to correctly
reproduce their effects from a knowledge of slowly varying configurations.
We will also illustrate directly the reconstruction of the \Sc functional for
rapidly varying fields from the local expansion.

We begin by formulating the \Sc equation for the vacuum functional and solving it to low order in `standard' perturbation theory, checking our result by computing the one-loop renormalisation of the coupling. We then
discuss the analyticity of the \Sc functional for complex scale 
parameters and use this to reformulate the \Sc equation so that it acts directly on a local expansion of the \Sc functional. We then solve this to low
order and show how the solution can be used to reconstruct the result of the
`standard' perturbative calculation. Finally we use a background field method to
calculate the beta-function from a local expansion.

\section{Semi-Classical expansion of the Yang-Mills Vacuum Functional}

In this section we will obtain the first few terms in the semiclassical
solution of the functional \Sc equation for the Yang-Mills vacuum functional, and check our results by computing the beta-function. This is the analogue of the usual loop expansion. It results in non-local expressions for the vacuum
functional because classical Yang-Mills theory is scale invariant, nonetheless
we will see that analyticity enables us to reconstruct these non-local terms from the local expansion that we obtain in the next section for the semiclassical expansion of the \Sc functional. 

\medskip
          The Yang-Mills action
can be written with the coupling factored out

\be S_{YM}=-{1\over g^2}\int d^3 x\, dt\,tr\left(\EE^2-\BB^2\right),\ee

\be \EE=-{\dot{\A}}+\nabla A_0+[\A,A_0],\quad \BB=\nabla\wedge\A+\A\wedge\A.\ee
(we denote Lie algebra generators of the
gauge group $SU(N)$ by $T^A$, so $\A=\A^CT^C,$ and $ (T^C)^\dagger=-T^C,\,tr\,(T^AT^B)=-
\delta^{AB}/2,\,[T^A,T^B]=f^{ABC}T^C$.) In the Weyl gauge ($A_0=0$) the canonical
coordinates are $\A^C({\bf x})$ and their conjugate momenta are
$-g^{-2}\EE^C({\bf x})$. The Hamiltonian is

\be H[\A,\EE]=  -{1\over g^2} \int d^3x\,tr\left( \EE^2+\BB^2\right). \ee
Since
$\dot  A_0({\bf x})$ does not appear in the action the Euler-Lagrange
equation obtained by varying $A_0({\bf x})$,

\be\nabla \cdot\EE+\A\cdot\EE-\EE\cdot\A=0,\ee
is a constraint rather
than a Hamiltonian equation of motion. In the
Schr\"odinger representation 
the field, $\BA(\xx)$, is diagonalised and
does not depend on time.
$\EE^C$ is represented by the operator

\be\EE^C ({\bf x})=ig^2 \hbar\frac{\del}{\del \BA^A (\xx)}  .\label{SR}
\ee
Substituting this into Gauss's law gives an operator. Requiring this to
annihilate physical wave-functionals

\be\Gamma \,\Psi[\A ]=0 ,\quad \Gamma\equiv \Big( \delta^{AB} 
\der_\alpha + f^{ACB} A_\alpha^C \Big) \frac{\del}{\del
A_\alpha^B} \ee
is simply the demand that these wave-functionals be gauge invariant 
because $\Gamma$ is the generator of time independent gauge transformations.
If $\tilde{\A}$ 
is the Fourier tranform of $\A$, 

\be
\BA^A (\xx) = \int \frac{d^3 p}{(2 \pi)^3}e^{i \, 
{\bf p} \,\, {\bf x}} \tilde {\BA}^A({\bf p}),\quad
\frac{\del A^A_\mu ({\bf p})}{\del A^B_\nu ({\bf q})} = (2 \pi )^3 \del^{AB} \del_{\mu
\nu} \del^3 ({\bf p}-{\bf q}),\ee
so that in momentum space Gauss's law becomes

\be
i p_\alpha \frac{\del \Psi}{\del A^A_\alpha (\pp)} +\int {d^3q\over (2\pi)^3} f^{ACB} A^C_\beta
(- \pp - \qq) \frac{\del \Psi}{\del A^B_\beta (- \qq)}=0 \, .
\label{GL}
\ee
The Hamiltonian operator that we would obtain by substituting the representation of the momenta, (\ref{SR}), into the classical expression
stands in need of regularisation because the two functional derivatives act at 
the same spatial point. We will regulate the Hamiltonian in a gauge-invariant
way so that it commutes with $\Gamma$ enabling us to simultaneously diagonalise both operators.
We will do this by inserting a gauge covariant kernel
between the two derivatives. This will depend on a short distance cut-off, $s$,
and we also need to introduce renormalisation constants. Symanzik \cite{SYMANZIK}-\cite{REV5} has shown that 
the \Sc equation exists for renormalisable field theories so that
the vacuum functional is finite as the cut-off is removed. For scalar field theories this
can involve the introduction of counter-terms associated with the quantisation surface
which are not seen in the renormalisation of usual Green's functions. 
For the case of Yang-Mills in $(3+1)$-dimensions gauge invariance precludes such additional terms, as discussed by
L\"{u}scher \cite{LUSCHER} and studied on the lattice by Sint et al \cite{SINT},  so that the physical states should be finite
if we simply make $g$ an appropriate function of the cut-off and subtract the vacuum energy density, $E(s)$. This gives the time-independent \Sc equation for an eigenstate
with energy $E$ above that of the vacuum in the form

\be
\lim_{s\downarrow 0}\left(
- \frac{\hbar^2g^2(s)}{2} \Delta_s   +  \frac{1}{2g^2(s)}{\cal B}-E-E(s) \right)\Psi =0
\label {SEQW}
\ee
where\be
\Delta_s = \int d^3x\,d^3y\frac{\del}{\del
A^R_\rho (\yy)} K^{RS}(\xx,\yy) \frac{\del }{\del A^S_\rho (\xx)} \, ,
\quad {\cal B}=\int d^3x \,\BB^R \cdot\BB^R\ee
There is considerable freedom in the choice of the kernel, $K$.
In this section we will choose $K$ to be a Wilson-line multipled by a momentum
cut-off, thus
\be
K^{RS}(\A,\xx, \yy) =U^{RS}(\xx, \yy) \int_{|\pp |^2 < 1/s}  \frac{d^3 p}{(2 \pi)^3}\, e^{i\, \pp . (\xx - \yy)}\ee
where
\be
 U^{RS}(\xx, \yy) = {\cal{P}} \, 
 exp \Big(  -\int_{\xx}^{\yy} A_\alpha^B({\bf{z}}) T^B  \, d z^\alpha \Big) \, ,
\label{WL}
\ee
${\cal{P}}$ being the usual path ordering operator, and the generators, $T^B$ are in the adjoint. $U^{RS}(\xx, \yy)$ is expandable as
\be
 U^{RS}(\xx, \yy) = \del^{RS} + {\cal{P}}\, 
\Bigg( \sum_{n=1} \frac{(-1)^n}{n!} \int_{\xx}^{\yy}
d {\mbox{z}}_{\mu_1} .. \int_{\xx}^{\yy} d {\mbox{z}}_{\mu_n} (A_{\mu_1}^{A^1} T^{A^1})
 ({\bf{z}}_1) .. (A_{\mu_n}^{A^n} T^{A^n}) ({\bf{z}}_n) \Bigg)_{RS} \, ,
\label{WLEXP}
\ee
and satisfies \footnote{For a proof, see \cite{Green} .}
\be
\frac{\del U^{RS} (\xx, \yy)}{\del A_\alpha^R (\yy)} = 0 \, .
\ee

      We will now consider the wave-functional of the vacuum. The functional integral representation that we discuss in the next section implies that for this
state the functional is of the form $\exp (W)$ where $W$ is perturbatively a sum of connected
Feynman diagrams. We expand $W$, $g^2(s)$ and $E(s)$ in powers of $\hbar$ as

\be
W[\BA]={1\over \hbar}\sum_{n=0}^\infty \hbar^nW_n[\BA]\,,\quad
g^2=\sum_{n=0}^\infty\hbar^n g^2_n \quad
E(s)=\sum_{n=1}^\infty\hbar^n E_n(s)
\ee
Order by order in $\hbar$, Gauss's law implies that each $W_n$ is gauge-invariant.
Substituting into the \Sc equation leads to
\bq
\lim_{s\downarrow 0}&&\Big\{
- \left(\Delta_s \sum_{n,m=0}^\infty \hbar^{n+m+1} g^2_n W_m  +
 \int d^3x\sum_{n,m} \hbar^{n+m+p}g^2_n\frac{\del W_n}{\del A^R_\alpha} 
\cdot \frac{\del W_m}{\del A^R_\alpha}\right)\nonumber\\
&&
 + {{\cal B}\over\sum_{n=0}^\infty\hbar^n g^2_n}  -2E(s)\Big\}=0
\label{SEQW22}
\eq
The order $\hbar^0$ or tree-level
contribution is the Hamilton-Jacobi equation for $W_0[\BA]$

\be
g_0^4\int {d^3p\over 2\pi^3} \frac{\del W_0}{\del A^R_\alpha (\pp)}
\frac{\del W_0}{\del A^R_\alpha (- \pp)} = {\cal
B} \, ,
\label{HJ}
\ee
which we will solve by further expanding 
in powers of $\BA$, so
\be
W_0=\sum_{n=2}^\infty W_0^n
\ee
where $W_0^n$ is of order $n$ in $\BA$, etc.
The next to leading order terms in the \Sc equation are of order
$\hbar$ 

\bq
\lim_{s\downarrow 0}&&\Big\{\left(g^2_0\Delta_s   W_0  +
g^2_1 \int d^3x \frac{\del W_0}{\del A^R} 
\cdot \frac{\del W_0}{\del A^R}+2g^2_0\int d^3x \frac{\del W_0}{\del A^R} 
\cdot \frac{\del W_1}{\del A^R}\right) 
\nonumber\\
&&+{g_1^2\over g_0^4} {\cal B} +2E_1(s)\Big\}
\nonumber \\
=\lim_{s\downarrow 0}&&\Big\{\left( g^2_0 \Delta_s W_0  
+2g^2_0\int d^3x \frac{\del W_0}{\del A^R} 
\cdot \frac{\del W_1}{\del A^R}\right) +{2g_1^2\over g_0^4} {\cal B} +2E_1(s)\Big\}=
0\label{onel}
\eq
Now $g_1^2$ is chosen to cancel the divergence coming from $\Delta_s   W_0$
that is proportional to $\cal B$. To calculate $g_1^2$, and hence obtain the one-loop
beta-function, we will have to first find $W_0$ to at least the fourth order
in $\BA$. In Appendix A we find the following expressions for
$ W_0^2$, $W_2^3$ and $W_0^4$:

\be
W_0^2 = {1\over g_0^2}\int {d^3p\over(2\pi)^3} A_\mu^A(\pp ) A_\nu^A( - \pp ) \,\, \Gamma^{\mu \nu} (\pp ),
\quad \Gamma_{\alpha \beta} ( \pp ) = 
\frac{p_\alpha p_\beta - \pp^2 \del_{\alpha \beta}}{2 |\pp |}.
\ee

\be
W_0^3 = {1\over g_0^2}\int {d^3p_1\,d^3p_2\over(2\pi)^6}
A_\alpha^A(\pp_1)A_\beta^B(\pp_2)A_\gamma^C(\pp_3)\,\,\, f^{ABC} \, \Gamma^{\alpha
\beta \gamma } (\pp_1,\pp_2,\pp_3) \,\, ,
\ee
with $\pp_3=-\pp_1-\pp_2$ and
\bq
&&\Gamma_{\alpha \beta \gamma}(\pp_1, \pp_2, \pp_3) = 
\nonumber \\&&
\frac{i }{3
(|\pp_1|+|\pp_2|+|\pp_3|)} \times {\Bigg\{} + \hat{p}_{1 \alpha} (\Gamma_{\beta \gamma}(\pp_3)  -  \Gamma_{\beta \gamma}(\pp_2) )
-\hat{p}_{2 \beta} (\Gamma_{\alpha \gamma}(\pp_3)  \nonumber \\
&-&   \Gamma_{\alpha \gamma}(\pp_1) )   
- \hat{p}_{3 \gamma} (\Gamma_{\alpha \beta}(\pp_1)  -  \Gamma_{\alpha \beta}(\pp_2) ) 
+ \frac{1}{2} ( p_{1 \eta} \delta^{\eta \alpha}_{\beta \gamma} +
                  p_{2 \eta} \delta^{\eta \beta}_{\gamma \alpha} +
                  p_{3 \eta} \delta^{\eta \gamma}_{\alpha \beta} )  
 {\Bigg\}}\,\,.
\label{G3}
\eq

\bq
W_0^4 &=& {1\over g_0^2}\int{d^3p_1\,d^3p_2\,d^3p_3\over(2\pi)^9}  \, A_\alpha^A (\pp_1) A_\beta^C (\pp_2) A_\gamma^B (\pp_3) A_\delta^D (\pp_4)
\nonumber \\
&&\quad\times f^{MAB}f^{MCD}\Gamma_{\alpha \beta \gamma \delta} (\pp_1, \pp_2, \pp_3, \pp_4) \, ,
\label{W42}
\eq
with $\pp_4=-\pp_1-\pp_2-\pp_3$ and

\bq
&& {\cal{S}}_4{\Big\{}f^{MAB}f^{MCD} \Gamma_{\alpha \beta \gamma \delta}
 (\pp_1, \pp_2, \pp_3, \pp_4){\Big\}} =\nonumber \\ && 
{\cal{S}}_4{\Big\{} f^{MAB}f^{MCD} \omega
\Big( 6i{\hat{p}}_{1 \alpha}  \Gamma_{\beta \gamma \delta}(\pp_1 +\pp_2,  \pp_3, \pp_4) 
 +\nonumber \\ && \,\,\,\,\,\,\,\,\,\,\,\
+9 \, \Gamma_{\eta \alpha
\beta}(\pp_3 + \pp_4, \pp_1, \pp_2)\Gamma_{\eta \gamma \delta}(\pp_1 + \pp_2, \pp_3,
\pp_4)-  {1\over 4}\del^{\alpha \beta}_{\gamma \delta} \Big){\Big\}}  
\label{G4}
\eq
where 
\be
\omega={1\over 2(|\pp_1| + |\pp_2| + |\pp_3| + |\pp_4|)}
\ee
and ${\cal S}_4$ symmetrises the expression under simultaneous interchanges of the four sets 
of indices and momenta $A,\alpha,\pp_1$ etc.

As a check on our calculation we now study the divergent behaviour of
$\Delta_s W_0$ as $s\downarrow 0$ and use this to calculate the beta-function.
This requires that we compute the divergent term quadratic in $\BA$ which
only receives contributions from $\Delta_s W_0^2$, $\Delta_s W_0^3$ and 
$\Delta_s W_0^4$. Because we have preserved gauge invariance this will have the form

\be
\xi \,\, ln \, s \, \int {d^3k\over (2\pi)^3} 
A_\mu^A({\bf{k}}) A_\nu^A (-{\bf{k}}) \Big(
k^\mu k^\nu - |{\bf{k}}|^2 \del^{\mu \nu} \Big)
\label{expected}
\ee
where $\xi $ is a constant. So to leading order in $s$

\be
\xi \,\, ln \, s \,  \Big(
k^\mu k^\nu - |{\bf{k}}|^2 \del^{\mu \nu} \Big)
=
\Lambda^{\mu\nu}_2+\Lambda^{\mu\nu}_3+\Lambda^{\mu\nu}_4\ee
where the $\Lambda^{\mu\nu}_r$ derive from $\Delta_s W_0^r$ and are given in Appendix B.
Taking the trace gives

\be
\xi \,\, ln \, s \,  2k^2
=
\Lambda^{\mu\mu}_2+\Lambda^{\mu\mu}_3+\Lambda^{\mu\mu}_4
\ee
From (\ref{C1}) we obtain
\bq
\Lambda_2^{\mu\mu}=&&
- 2 N\,  \int_{\pp^2 < 1/s} {d^3p\, d\sigma_2\, dy\over (2\pi)^3} \, \frac{1}{|\pp - {\bf{k}} y|} \nonumber \\
&=& -4\pi N\int_{ \pp^2 < 1/s} d\sigma_2\, dy\pp^2 d|\pp | \, d\theta_p \,
\frac{\sin \theta_p}{(\pp^2 - 2 |\pp ||{\bf{k}}| \cos \theta_p y + {\bf{k}}^2
y^2)^{1/2}} \, .
\label{tr-C1}
\eq
Furthermore, in order to compute its contribution to (\ref{expected}), we need only
the ${\bf{k}}^2/\pp^3$ term of the expansion of (\ref{tr-C1}) for large $|\pp|$. After 
performing the remaining integrations, we conclude that there is no contribution
from $\Delta_s W_0^2$ to (\ref{expected}).  The same holds for $\Delta_s W_0^3$, as
we can verify by taking the trace of (\ref{C2}) and expanding for large $|\pp|$.
However, if we take the trace of (\ref{C3}) integrate over $\theta_p$
and expand for large $|\pp|$ we get
\be
\Lambda_4^{\mu\mu}=
\frac{  N }{(2 \pi)^2}
 \int_{|\pp|^2<1/s}
|\pp |^2 \, d|\pp | \, \frac{222}{3} \frac{{\bf{k}}^2}{\pp^3}\,,
\ee
hence $\xi=11N/(24\pi^2)$. For the divergences to cancel in
(\ref{onel}) we need to take $2g_1^2/ g_0^4=\xi \log (s\mu^2)$
where $\mu$ is a mass scale. So to one loop the bare coupling is
\be
g^2=g_0^2+\hbar g_1^2=g_0^2+\hbar{11g^4_0N\log (s\mu^2)\over 48\pi^2}\, ,
\ee
giving the usual result for the one-loop beta-function
\be\beta=\mu{\partial g^2\over \partial\mu}=-\hbar{11g^4_0N\over 24\pi^2}\, .
\ee


\section{The local expansion of the \Sc functional}
In this section we will argue that the wave-functionals of the states in Yang-Mills theory can be reconstructed from a local expansion that satisfies its own form of the \Sc equation.
This reduces to an infinite set of coupled {\it algebraic} equations that are amenable to numerical solution without recourse to perturbation theory,
as in \cite{forth}, but our purpose here is to 
show that these are consistent with the known methods of perturbation theory where applicable. In particular we will show that the expressions of the previous section, and in particular the calculation of the beta-function, can be reproduced using the local expansion.
Because Yang-Mills theory in 3+1 dimensions is scale-invariant the 
semiclassical expansion of the vacuum functional for which we have obtained the first few terms
is non-local. For example, the lowest order contribution, $W_0^2$ in configuration space 
is proportional to

\be
\int d^3x\, d^3y\,\nabla\wedge\A^C ({\bf x}) \cdot \nabla\wedge\A^C ({\bf y})/({\bf x}-{\bf y})^2
\ee
Asymptotic freedom implies that such expressions are appropriate when $\A$ is a rapidly varying function of ${\bf x}$
on the scale of the  masses of the physical particles in the theory, the glueballs,
so the one-loop beta-function results from high momentum modes of the field.
The local expansion that we will use, being an expansion in powers of derivatives is appropriate to low momenta. To reconstruct the high momentum
behaviour from that at low momentum we show that if we perform a scale-transformation on the 
wave-functionals with a real scale parameter $\rho$ then the wave-functional
becomes an analytic function in some domain when we allow $\rho$ to be complex. This will enable us to use Cauchy's theorem to
relate the dependence of the scaled wave-functional for large $\rho$ to that at small
$\rho$. 

         Rather than deal with the vacuum functional we will study the \Sc functional which is
the matrix element of the Euclidean time evolution operator in a basis in which 
the gauge-potential has been diagonalised.
This has two advantages. Firstly, from $\Phi$ we can extract the energy eigenstates
and eigenvalues by inserting a basis of such states 

\be
\langle {\bf A }|  e^{-{\em H}  \tau /\hbar} | {\bf A}' \rangle 
=\sum_{E}\Psi_{E} [\A] \Psi_{E}^*[\A'] e^{-\tau
E} 
\ee
At large times this is dominated by the contribution from the vacuum.
Secondly the Euclidean time $\tau$ plays the r\^ole of an inverse mass enabling us to
construct a derivative expansion in positive powers of $\tau$ which at the same time we can
study within the semiclassical expansion. 
Although such an expansion will only converge for small $\tau$
we will be able to use it to reconstruct the \Sc functional for arbitrary
fields and once this is done $\tau$ can take any value (consistent with 
an approximation that we will later describe).
We will now show that if we scale $\A$ and
$\A'$ then $\Phi$ becomes an analytic function of the scaling parameter.
This generalises the discussion of the analyticity of the vacuum functional 
given in \cite{PM}. 


\subsection{Analyticity of the \Sc Functional}

Symanzik showed how to construct the \Sc functional for scalar field theory in terms of a  functional integral over a spacetime bounded by two parallel planes, by placing sources 
on the planes coupled to the normal derivative of the quantum field,
\cite{SYMANZIK}-\cite{REV5}.
We will generalise this to take account of gauge invariance
by representing the \Sc functional as 
\be
\Phi_{\tau}[{\bf A}, {\bf A}']=\int{\cal D}A\,e^{-S[A]-S_b[A,\, {\bf A},\,{\bf A}']}
\label{eq:funct}
\ee
where $S[A]$ is the Yang-Mills action gauge-fixed in the 
gauge $A_0=0$. Space-time is Euclidean with co-ordinates 
$({\bf x},t)$ and $0\le t\le \tau$, so
\be
S[A]=-{\textstyle {1\over {\tilde g}^2}}\int d^3{\bf x}\,dt\,tr\,\left(
{\dot A}^2 + (\nabla \wedge{ A}+{ A}\wedge { A})^2\right)
\ee
The boundary term in 
the action is 
\be
S_b[A,{\bf A},\,{\bf A}']]=-{\textstyle {2\over g^2}}\int d^3{\bf x}\,tr\,\left(
({\bf A}({\bf x})-A({\bf x},\tau))\cdot{\dot A}({\bf x},\tau)
-({\bf A}'({\bf x})-A({\bf x},0))\cdot{\dot A}({\bf x},0)
\right)
\ee
The boundary values of $A$ are to be freely integrated over.
We will assume that at spatial infinity the sources $\bf A$ and 
${\bf A}'$ are 
pure gauges ${\bf A},\BA'\sim g(\hat{\bf x})^{-1}\nabla g(\hat{\bf x})$.
$S_b$ is chosen so that $\Phi_\tau[{\bf A},{\bf A}']$ is invariant under the 
time-independent gauge
transformation acting simultaneously on the sources 
$\delta_\omega {\bf A}=\nabla\omega +[{\bf A},\omega]$, 
and
$\delta_\omega {\bf A}'=\nabla\omega +[{\bf A}',\omega]$
since the effect of varying the sources may be compensated by 
gauge transforming $A$. As $\omega$ cannot depend on time this is
the residual gauge symmetry of $S[A]$ that preserves the gauge 
condition. We will assume that there is a regulator in place that 
preserves this invariance and on which the 
couplings $g,\tilde g$ depend. 
Functionally differentiating with respect to the source $\BA$
leads to an insertion of ${\dot A}$ on the boundary $t=\tau$.
Given the Wick rotation between Euclidean time, $t$,
and Minkowskian time, this yields the \Sc representation
of the canonical momentum.

We now study the effect of a scale-transformation of the sources.
Consider scaling each dimension separately and set
$\A^{\rho}=(\rho_1\rho_2\rho_3)^{-1/6}\A (x^1/\sqrt \rho_1, x^2/\sqrt \rho_2,
x^3/\sqrt \rho_3)$ and $\A'^{\rho}=(\rho_1\rho_2\rho_3)^{-1/6}\A' (x^1/\sqrt \rho_1, x^2/\sqrt \rho_2,
x^3/\sqrt \rho_3)$. We will show that $\Phi_\tau [\A ^{\rho},\A '^{\rho}]$ is analytic 
in $\rho_1,\rho_2,\rho_3$ separately.
Firstly
we interchange the names of the Euclidean time, $t$, and one of the
spatial co-ordinates, $x^1$ say, in the functional integral 
(\ref{eq:funct})
which we now interpret as the Euclidean time-ordered vacuum 
expectation value
\be
\Phi_{\tau}[{\bf A}^\rho, {\bf A}'^\rho]=T\langle 0_r|\,exp\,\left({\textstyle 
{2\over g^2}}\int dx^2\,dx^3\,
dt\,tr\,\left( \A^{\rho}\cdot \hat A^\prime|_{x^1=\tau}-\A'^{\rho}\cdot \hat A^\prime|_{x^1=0}\right)\right)\,
|0_r\rangle
\ee
where $|0_r\rangle$ is the vacuum for the Yang-Mills Hamiltonian 
defined on the
space $0\le x^1\le \tau$ in the axial gauge $A_1=0$ with  boundary 
terms in the
action $\pm {\textstyle 1\over g^2}\int dx^2\,dx^3\,
dt\,tr\,\left( A\cdot \hat A^\prime\right)$, $\hat A^\prime$ is
the derivative of $\hat A$ with respect to $x^1$, and $\hat A (x_1,x_2,x_3,t)$ is the field operator. 
To make the notation more compact introduce ${\cal A}_z$ and $\tilde A_z$
where the index $z$ takes the two values $\pm$ and
\bq
{\cal A}_+=\BA,\quad && {\cal A}_-=-\BA',
\nn
{\tilde  A}_+(x_2,x_3,t)=\hat A (\tau,x_2,x_3,t),
\quad && {\tilde A}_-(x_2,x_3,t)=\hat A(0,x_2,x_3,t),
\label{defcalA}
\eq
with similar relations for the scaled variables.
Then the coupling of the sources to the quantum field becomes
 \be
\A^{\rho}\cdot \hat A^\prime|_{x^1=\tau}-\A'^{\rho}\cdot \hat A^\prime|_{x^1=0}
={\cal A}^\rho_z\cdot{\tilde A_z}\, .
\ee

Expanding the exponential and using the time-evolution operator for the
rotated theory, $\exp (-\hat H_r t)$, to generate the t-dependence of
the quantum fields
gives
\bq
&&\Phi_{\tau}[{\bf A}^\rho, {\bf A}'^\rho]=
\nn
&&\sum_n\int_{-\infty}^\infty dt_n   \int_{-\infty}^{t_{n-1}} 
dt_{n-1}
..\int_{-\infty}^{t_3} dt_2\int_{-\infty}^{t_2} dt_1 
\prod_{i=1}^n\left({\textstyle {-{1\over g^2}}}\int dx^2_i\,dx^3_i
\,{\cal A}^\rho_{R_i}
(t_i,x_i^2,x^3_i)\right)
\nn
&& \times
\langle 0_r|{\tilde A}^\prime _{R_n}(x^2_n,x^3_n,0)\,e^{(t_{n-1}-t_n){\hat H}_r}
..
{\tilde A}^\prime_{R_2}(x_2^2,x_3^3,0)\,
e^{(t_2-t_1){\hat H}_r}\,{\tilde A}^\prime_{R_1}(0,x_1^2,x_1^3)\,|0_r\rangle.
\label{eq:expexp}
\eq
Here $R_i$ stands for all the indices, i.e. Lie algebra and spatial indices
and $z$.
The time integrals may be done after Fourier transforming the sources.
To do this we define the $\rho_1$-independent Fourier mode
\be
a(k,x^2,x^3)\equiv
(\rho_2\rho_3)^{-1/6}\int dt\,e^{-ikt}\,{\cal A }(t,x^2/\sqrt \rho_2 ,
x^3/\sqrt \rho_3),
\ee
so that
\be
{\cal A} ^{\rho} (t,x^2,x^3)={\textstyle {1\over 2\pi}}\int dk \, 
e^{ikt/\sqrt \rho_1}\,
\rho_1^{-1/6}\,a(k,x^2,x^3).
\ee
Substituting this into (\ref{eq:expexp}) and inserting
the resolution of the identity in terms of eigenstates of $H$ 
gives
\bq
&&\Phi_{\tau}[{\bf A}^\rho, {\bf A}'^\rho]=
\nn
&&\sum_n 
\prod_{i=1}^n\left({\textstyle {-{1\over \pi g^2}}}\int
dk_i\, dx^2_i\,dx^3_i\,\rho_1^{1/3}\, a_{R_i}
(k_i,x_i^2,x^3_i)\right)\,\delta\left( \sum k_i \right)
\nn
&&\sum_{E_1..E_{n-1}}
\langle 0_r|A^\prime _{R_n}(x^2_n,x^3_n,0)\,|E_{n-1}\rangle {1\over 
\sqrt \rho_1 E_{n-1}-i\sum_1^{n-1}k_j}
..\nn
&&\quad..
{1\over \sqrt \rho_1 E_1-ik_1}\langle E_1|\,A^\prime_{R_1}(x_1^2,x_1^3,0)\,
|0_r\rangle.
\label{expa}
\eq
This makes explicit the $\rho_1$-dependence of $\Phi_{\tau}[{\bf A}^\rho, {\bf A}'^\rho]$. 
Although $\rho_1$ was originally real and positive we may use this
expression to define an analytic continuation to complex values, 
yielding
a function that is analytic away from the zeroes of the 
denominators in
(\ref{expa}). Since the eigenvalues of the Hermitian 
Hamiltonian are real,
the singularities lie on the negative real $\rho_1$-axis. 
Similarly we may show 
that $\Phi_{\tau}[{\bf A}^\rho, {\bf A}'^\rho]$ 
 continues to an analytic function in 
$\rho_2$ and $\rho_3$
and, by setting $\rho_1=\rho_2=\rho_3=\rho$, that $\Phi_{\tau}[{\bf A}^\rho, {\bf A}'^\rho]$ 
 continues to 
an analytic function in $\rho$ in the complex plane with the 
negative real axis 
removed. This scaling preserves gauge invariance.

    We can see that (\ref{expa}) has a derivative expansion provided the
source terms have compact support in momentum space.
Consider the logarithm of $\Phi_\tau$. As a series in powers of the sources this can be written as sums of products of terms of the same form as appear on the
right-hand-side of (\ref{expa}). So any term of a definite order in the
sources, being a finite sum of analytic functions will also be analytic.
Yang-Mills theory has a mass-gap so, apart from the contribution of the vacuum, 
the eigenvalues $E_n$ are all greater than zero, and the energy denominators can be expanded in positive
powers of momenta, $k_i$, when $\rho$ is sufficently large. 
When one of the intermediate states in the sum is the vacuum the 
energy denonminator is just $1/(-i\sum_1^{n-1}k_j)$ which does not have 
such an expansion. However such denominators
must cancel against powers of momentum in the numerator
so as not to violate the cluster decomposition property of the logarithm of $\Phi_\tau$.
This will be the case within the semi-classical expansion even though the glueball
mass cannot be seen in perturbation theory. This is because the Hamiltonian $H$ 
describes a theory defined on the space $0<x_1<\tau$, so the masses of
particles will be greater than $\approx 1/\tau$.
So for large $\rho$ the logarithm of $\Phi_\tau[{\bf A}^\rho, {\bf A}'^\rho]$ looks like

\bq 
&&log(\Phi_{\tau}[{\bf A}^\rho, {\bf A}'^\rho])=\nn
&&
W[{\bf A}^\rho,\BA'^\rho]=\int d^3{\bf x}\,
(a_1\rho^{-1/2}tr\,{\bf B}\cdot{\bf B} +a_2\rho^{-3/2}tr\,D
\wedge{\bf B}\cdot
D\wedge{\bf B}
\nonumber\\
& &
+a_3\rho^{-3/2}tr\,{\bf B}\cdot
({\bf B}\wedge{\bf B}) +..a_1\rho^{-1/2}tr\,{\bf B}'\cdot{\bf B}' +a_2\rho^{-3/2}tr\,D'
\wedge{\bf B'}\cdot
D'\wedge{\bf B}'
\nonumber\\
& &
+a_3\rho^{-3/2}tr\,{\bf B}'\cdot
({\bf B}'\wedge{\bf B}') +..
{\bar a_1}\sqrt \rho tr\, (\BA-\BA')^2)+..)
\label{YMexpp}
\eq
where the coefficents $a_n$ and $\bar a_n$ depend on $\tau$, but not on $\rho$ or on $\bf x$.
Note that parity invariance requires that the powers of $\rho$ are all
half-odd-integers. Thus the large $\rho$ expansion of $\sqrt\rho\, W[{\bf A}^\rho,\BA'^\rho]$
has only integer powers of $\rho$ indicating that this quantity is analytic in the complex 
$\rho$-plane with a cut on the negative real axis of finite length running from the origin. 
We can now use Cauchy's theorem to express $W[{\bf A},\BA']$ in terms of this expansion,
so that the \Sc functional for arbitrary sources $\BA,\BA'$ can be reconstructed 
from a knowledge of its value when the  sources vary slowly and for which the local expansion is
valid. We will evaluate the contour integral 
\be
I(\lambda)={1\over 2\pi i}\int_C {d\rho\over \rho-1}\,e^{\lambda (\rho-1)}
\, \sqrt\rho\, W[{\bf A}^\rho,\BA'^\rho]\label{I}
\ee
in two ways. We take $C$ to be a circle centred on the origin and large enough for us to
be able to use the local expansion. Re-expanding in inverse powers of $\rho-1$ and writing the result as
$\sqrt\rho\, W[{\bf A}^\rho,\BA'^\rho]= \sum \,(\rho-1)^{-n}\omega_n[\A ,\A']$
enables us to calculate

\be
I(\lambda)=\sum_n {\lambda^n \omega[\A ,\A']_n\over \Gamma (n+1)}
\ee

The integral may also be evaluated by collapsing the contour
$C$ until it breaks into two disconnected pieces, a small circle 
centred on
$\rho=1$ and a contour that
surrounds the finite cut on the negative real axis. The integral over the 
circle 
gives $W[{\bf A},\BA']$, so

\be
I(\lambda)=\int _0^adx\,(e^{-\lambda (x+1+i\epsilon)}F_+(x)
-e^{-\lambda (x+1-i\epsilon)}F_-(x)) +W[\BA,\BA']
\ee
where we have taken the contour surrounding the cut to consist of two line segments of length $a$ parallel to the real-axis a distance $\epsilon$ from it,
and 
\be
F_\pm=\sqrt\rho\, W[{\bf A}^\rho,\BA'^\rho]/(2\pi i(1-\rho))\quad{\rm for}\quad\rho=-x\pm i\epsilon .
\ee
The contribution from the semi-cicle
needed to close the contour near the origin is negligible, being of
order $\epsilon^{3/2}$,
because asymptotic freedom implies that for small $\rho$ the functional $W[{\bf A}^\rho,\BA'^\rho]$
is reliably given by its tree-level value which is independent of 
$\rho$, \cite{PM}. If we set $\lambda=u+iv$, and consider $u>0$,
the magnitude of the integral is bounded 
\be
|\int _0^adx\,(e^{-\lambda (x+1+i\epsilon)}F_+(x)
-e^{-\lambda (x+1-i\epsilon)}F_-(x))|
< a|F_+|^{max}e^{-u+v\epsilon}+a|F_-|^{max}e^{-u-v\epsilon}
\ee
where $|F_\pm|^{max}$ is the maximum value taken by $|F_\pm|$
on the contour.
So the contribution from the cut goes to zero as $u\rightarrow \infty$.
Thus we reach our goal of being able 
to reconstruct the \Sc functional from the local expansion.

\be 
W[{\bf A},\BA']=
\lim_{\lambda\rightarrow\infty}
\sum_n {\lambda^n \omega[\A ,\A']_n\over \Gamma (n+1)}
\label{eeee}
\ee

Note that the denominators in (\ref{eeee}) improve the convergence of the
series compared to the original series representing the large
$\rho$ behaviour of $\sqrt\rho\, W[{\bf A}^\rho,\BA'^\rho]$ from
which it was constucted. We can improve this convergence still further
since $I(\lambda)$ is analytic and bounded in the region
$u>|v|\epsilon$, so if we set $\rho'=1/\lambda^2$ then
$I( 1/\sqrt\rho')$ is analytic in the complex $\rho'$ plane with a
small wedge surrounding the negative real axis removed. Since it is bounded throughout this region we may repeat the above argument to 
obtain
\be
\lim_{\lambda\rightarrow\infty}I(\lambda)=
\lim_{\lambda\rightarrow\infty}{1\over 2\pi i}\int_{C'} {d\rho'\over \rho'}\,e^{\lambda^2 \rho'}
I( 1/\sqrt\rho')
\ee
where $C'$ is that part of a circle of very large radius centred on the origin
that is outside the wedge. Let us denote the operation of applying the integrals
over $\rho$ and $\rho'$ to $\sqrt\rho\, W[{\bf A}^\rho,\BA'^\rho]$
by $R_\lambda$, i.e.

\be
R_\lambda \sqrt\rho\, W[{\bf A}^\rho,\BA'^\rho]=
{1\over 2\pi i}\int_{C'} {d\rho'\over \rho'}\,e^{\lambda^2 \rho'}
{1\over 2\pi i}\int_C {d\rho\over \rho-1}\,e^{(\rho-1)/\sqrt\rho'}
\, \sqrt\rho\, W[{\bf A}^\rho,\BA'^\rho]
\ee
Applying this to the series (\ref{eeee}) gives (if we take $C'$ to be a
complete circle)
\be 
W[{\bf A},\BA']=\lim_{\lambda\rightarrow\infty}R_\lambda \sqrt\rho\, W[{\bf A}^\rho,\BA'^\rho]=
\lim_{\lambda\rightarrow\infty}
\sum_n {\lambda^n \omega[\A ,\A']_n\over \Gamma (n+1)\Gamma(n/2+1)}
\label{eeeee}
\ee
Although this requires taking a limit of an infinite series of positive powers of
$\lambda$ we will see that a good 
approximation is obtained by truncating the series to
a finite number of terms and taking $\lambda$ as large
as is consistent with this truncation.

\subsection{The \Sc equation for the local expansion}

Symanzik's work implies that the logarithm of the \Sc functional
satisfies the \Sc equation which we write as

\bq &&\lim_{s\downarrow 0}\quad\Big (-{\hbar\over g^2(s)}{\partial W[{\bf A},\BA']\over\partial\tau}+
\nn
&&
\frac{\hbar^2}{2}\left( \Delta_sW[{\bf A},\BA']+ \int d^3x \frac{\del W[{\bf A},\BA']}{\del A^R} 
\cdot \frac{\del W[{\bf A},\BA']}{\del A^R}\right)  -  \frac{1}{2g^4(s)}{\cal B}+{E(s)\over g^2(s)} \Big)=0
\label{SEQSF}
\eq
This determines the coefficents $a_1,a_2,..\bar a_1,..$ of the local expansion but we cannot simply substitute (\ref{YMexpp}) into (\ref{SEQSF}) since the local expansion is only valid for
slowly varying fields whereas the $s\downarrow 0$ limit requires a knowledge 
of high momentum modes. Nonetheless we can find a version of the \Sc equation that is
satisfied by the local expansion by considering the analyticity of $\Delta_sW$ as a function of
a scaled cut-off. If we write $W[{\bf A},\BA']$ as
\be
W[{\bf A},\BA']=\sum_n\int(\prod_{i=1}^n d^3x_i\, {\cal A}_{R_i}({\bf x}_i))\,
\Gamma_n^{R_1..R_n}({\bf x}_1,..{\bf x}_n)
\ee
(with ${\cal A}$ defined by (\ref{defcalA})) 
then we have shown that $\Gamma_n^{R_1..R_n}(\sqrt\rho{\bf x}_1,..\sqrt\rho{\bf x}_n)$ is analytic
in the cut $\rho$-plane. Applying the functional Laplacian to this expression gives

\bq
(\Delta_{s}W)[\BA,\BA']&=&\sum_nn(n-1)\int(\prod_{i=3}^n d^3x_i\, {\cal A}_{R_i}({\bf x}_i))\,\int d^3x\,d^3y\times\nn
&& K^{RS}_{s}(\A,\xx,\yy) \,\Gamma_n^{(R\nu,+) \,(S\nu+)\, R_3..R_n}(\xx,\yy,{\bf x}_1,..{\bf x}_n)\eq
So if we calculate $\Delta_s W$ evaluated for the scaled gauge-potential $\BA^\rho$ and for a cut-off $s=\rho/\mu^2$, with $\mu$ a mass-scale,
then
\bq
(\Delta_{\rho/\mu^2}W)[\BA^\rho,\BA'^\rho]&=&\sum_nn(n-1)\int(\prod_{i=3}^n d^3x_i\, {\cal A}^\rho_{R_i}({\bf x}_i))\,\int d^3x\,d^3y\times\nn
&& K^{RS}_{\rho/\mu^2}(\A^\rho,\xx,\yy) \,\Gamma_n^{{\hat R_1} {\hat R_2} R_3..R_n}(\xx,\yy,{\bf x}_1,..{\bf x}_n),\label{der}\eq
where ${\hat R_1}$ stands for the set of indices $\{R,\nu,+\}$ 
and  ${\hat R_2}$ stands for $\{S,\nu,+\}$. 
Now
\bq
&&K^{RS}_{\rho/\mu^2}(\A^\rho,\xx, \yy) ={\cal{P}} \, 
 exp \Big(  -\int_{\xx}^{\yy} A_\alpha^B({\bf{z}/\sqrt\rho}) T^B  \, {d z^\alpha/\sqrt\rho} \Big) \int_{|\pp |^2 < \mu^2/\rho}  \frac{d^3 p}{(2 \pi)^3}\, e^{i\, \pp . (\xx - \yy)}\nn
&&\quad\quad={\cal{P}} \, 
 exp \Big(  -\int_{\xx/\sqrt\rho}^{\yy/\sqrt\rho} A_\alpha^B({\bf{z}}) T^B  \, {d z^\alpha} \Big) \int_{|\pp |^2 < \mu^2}  \frac{d^3 p}{(2 \pi{\sqrt\rho})^3}\, e^{i\, \pp . (\xx/{\sqrt\rho} - \yy/{\sqrt\rho})}
\eq
so changing the integration variables in (\ref{der}) gives

\bq &&\sum_nn(n-1)\int(\prod_{i=3}^n  d^3x_i\, {\cal A}_{R_i}({\bf x}_i))\,\int d^3x\,d^3y\,\,\times\nn
&& K^{RS}_{1/\mu^2}(\A,\xx,\yy) \,\rho^{n-1/2}\Gamma_n^{{\hat R_1} {\hat R_2} R_3..R_n}(\sqrt\rho\xx,\sqrt\rho\yy,\sqrt\rho{\bf x}_1,..\sqrt\rho{\bf x}_n).
\eq
All the $\rho$ dependence is now contained in the $\rho^{n-1/2}\Gamma_n$ factors, and these are analytic
in the cut $\rho$-plane. For large values of $\rho$ we can calculate this by simply applying 
the  Laplacian to the  
local expansion (\ref{YMexpp}) which yields a power series in integer powers of $\rho$,
showing that the cut along the negative $\rho$-axis is of finite length for
any term of fixed order in the sources. Following the same argument as earlier 
we will be able to obtain the small $s$ behaviour of $\Delta_sW[{\bf A},\BA']$ from that for large $\lambda$ by employing Cauchy's theorem, but it is advantageous  to do this for all the 
terms in the \Sc equation (\ref{SEQSF}) so we now turn our attention to the renormalisation constants. 

The $s$-dependence of $g(s)$ and $E(s)$ for small $s$ are chosen to make $W$ finite as the cut-off is removed. We are free to choose their values for $s$ away from
$0$ provided they are consistent with this small $s$ behaviour.
We will now show that the other terms in the \Sc equation, namely $E(s)/g^2(s)$
and $1/g^4(s)$ can be arranged to be analytic in $\rho$ when $s=\rho/\mu^2$.   The \Sc equation (\ref{SEQSF}) holds for all sources $\BA$ and $\BA'$.
If we evaluate it for sources that vary slowly over distances of order $\tau$ then both sides are sums of local
expressions and we can unambiguously extract the coefficent of $\cal B$, and the source
independent term, so

\bq &&\lim_{s\downarrow 0}\quad\Big({{\hbar\over 2g^2(s)}}{\partial a_1\over\partial\tau}+
\nn
&&
\frac{\hbar^2}{2}\left( \Delta_sW[{\bf A},\BA']+ \int d^3x \frac{\del W[{\bf A},\BA']}{\del A^R} 
\cdot \frac{\del W[{\bf A},\BA']}{\del A^R}\right){\Big |}_{\cal B} -{1\over 2g^4(s)}\Big)=0
\label{SEQSFR}
\eq
and 
\be 0=
\lim_{s\downarrow 0}\left(
 \frac{\hbar^2}{2}\left( \Delta_sW\right)[0,0]  +{E(s)\over g^2(s)} \right)
\label{SEQSFR2}
\ee
These equations hold for all time $\tau$, and in particular for the limit of very large 
times when they are dominated by contributions from the vacuum. In this case the $\tau$-derivative of $a_1$ tends to zero as $\tau$ becomes large. We will now {\it choose} the $s$-dependence of $g(s)$ and $E(s)$ 
so that 
\be {1\over g^4(s)}
=
\lim_{\tau\rightarrow \infty}
 \hbar^2\left(  \Delta_sW[{\bf A},\BA']+ \int d^3x \frac{\del W[{\bf A},\BA']}{\del A^R} 
\cdot \frac{\del W[{\bf A},\BA']}{\del A^R}\right){\Big |}_{\cal B} \label{SEQSFR3}
\ee
and
\be 0=
\lim_{\tau\rightarrow \infty}\left(
 \frac{\hbar^2}{2}\left( \Delta_sW\right)[0,0]  +{E(s)\over g^2(s)} \right)
\label{SEQSFR4}
\ee
This choice clearly satisfies (\ref{SEQSFR3}) and (\ref{SEQSFR4}) and when we set 
$s=\rho/\mu^2$ gives $E(s)/g^2(s)$ and $1/g^4(s)$ as analytic functions of
$\rho$ in the cut $\rho$-plane, with the cuts on the negative real axis having finite length. This implies that $1/g^2(s)$ has the same property
provided $1/g^4(s)$ as defined by (\ref{SEQSFR3}) has no  zeroes for finite
$\rho$ away from the negative real axis. (If this is not the case we can
still define such an analytic function that for small $s$ agrees with $1/g^2(s)$ by demanding that for a particular value of $\tau$ (\ref{SEQSFR2}) hold for all $s$, although this would not necessarily square to (\ref{SEQSFR3})).
This enables us to obtain the limit as $s\downarrow 0$ that occurs in the \Sc equation (\ref{SEQSF}) as

\bq &&\lim_{\lambda\rightarrow\infty}{1\over 2\pi i}\int_C {d\rho\over \rho}\,e^{\lambda \rho}\Big\{-
\,{\hbar\over g^2(\rho/\mu^2)}{\partial W[{\bf A}^\rho,\BA'^\rho]\over\partial\tau}
\nn
&&
+\, \frac{\hbar^2}{2}\left (
\Delta_{\rho/\mu^2}W[{\bf A}^\rho,\BA'^\rho]+\int d^3x \frac{\del W[{\bf A},\BA']}{\del A^{ R}} \cdot \frac{\del W[{\bf A},\BA']}{\del A^{ R}} \right)
  \nn &&\quad\quad-  \frac{1}{2g^4(\rho/\mu^2)}{\cal B}^\rho+{E(\rho/\mu^2)\over g^2(\rho/\mu^2)} 
\Big\}=0.
\label{SEQSF2'}
\eq
As before $C$ is a circle centred on the origin and of very large radius so that we can
calculate using the local expansion. Since both sides are analytic in the cut $\rho$-plane we can
collapse the contour to a component surrounding the cut and a circle centred on the origin of very small radius whose contribution to (\ref{SEQSF2'})
 is just the original \Sc equation,
and the contribution from the cut is damped by the $e^{\lambda\rho}$ factor, since for any 
term in the local expansion the cut has a finite length, and so the magnitudes of the
other terms in the integrand are bounded. In practice we will use 
series expansions to express the large $\rho$-behaviour. The $\rho$-integration then leads to $\Gamma$ functions that improve the convergence of the series.
As before this convergence can be improved still further by an additional integration. Thus if we denote by $R'_\lambda$ the operator

\be
R_\lambda'  f(\rho)=
{1\over 2\pi i}\int_{C'} {d\rho'\over \rho'}\,e^{\lambda^2 \rho'}
{1\over 2\pi i}\int_C {d\rho\over \rho}\,e^{\rho/\sqrt\rho'}
\,  f(\rho)
\ee
then

\bq &&\lim_{\lambda\rightarrow\infty}R_\lambda'
\Big\{-
\,{\hbar\over g^2(\rho/\mu^2)}{\partial W[{\bf A}^\rho,\BA'^\rho]\over\partial\tau}
\nn
&&
+\, \frac{\hbar^2}{2}\left (
\Delta_{\rho/\mu^2}W[{\bf A}^\rho,\BA'^\rho]+\int d^3x \frac{\del W[{\bf A}^\rho,\BA'^\rho]}{\del A^{ R}} \cdot \frac{\del W[{\bf A}^\rho,\BA'^\rho]}{\del A^{ R}} \right)
  \nn &&\quad\quad-  \frac{1}{2g^4(\rho/\mu^2)}{\cal B}^\rho+{E(\rho/\mu^2)\over g^2(\rho/\mu^2)} 
\Big\}=0.
\label{SEQSF2}
\eq

\subsection{Solving the \Sc equation for the local expansion in powers of $\hbar$}

Substituting the local expansion (\ref{YMexpp})
into (\ref{SEQSF2}) leads to algebraic
equations for the coefficents which can be solved 
without recourse to perturbation theory, as has been done for $\phi^4$-theory
\cite{forth}. 
However our purpose here is limited to showing that the local expansion reproduces
semiclassical perturbation theory and leads to the correct result for the one-loop
beta-function, even though this depends on large momentum modes and the local expansion appears to only contain information about small momentum. As in the case of the vacuum functional we expand

\be
W[\BA,\BA']={1\over \hbar}\sum_{n=0}^\infty \hbar^nW_n[\BA,\BA']\,,\quad
g^2=\sum_{n=0}^\infty\hbar^n g^2_n \,,\quad
E(s)=\sum_{n=1}^\infty\hbar^n E_n(s)
\ee
The tree-level
contribution is 
\bq &&0=\lim_{\lambda\rightarrow\infty}R_\lambda '\,\Big({1\over g_0^2(\rho/\mu^2)}{\partial W_0[{\bf A^\rho},\BA'^\rho]\over\partial\tau}
\nn
&&
- \frac{1}{2}\int d^3x \frac{\del W_0[{\bf A^\rho},\BA'^\rho]}{\del A^{ R}} 
\cdot \frac{\del W_0[{\bf A}^\rho,\BA'^\rho]}{\del A^{ R}}  
+  \frac{1}{2g_0^4(\rho/\mu^2)}{\cal B}^\rho\Big).
\label{SEQSFT}
\eq
The absence $\Delta W[{\bf A},\BA']$ means that once the local expansion has been substituted into this equation the $\rho$-dependence of each term is given by its engineering 
dimension. The condition (\ref{SEQSFR3})
to this order in $\hbar$
is
\be 0=
\lim_{\tau\rightarrow \infty}\left(
 \frac{\hbar^2}{2}\left( \int d^3x \frac{\del W[{\bf A},\BA']}{\del A^R} 
\cdot \frac{\del W[{\bf A},\BA']}{\del A^R}\right){\Big |}_{\cal B} -{1\over 2g_0^4(s)}\right)
\label{SEQSFR42}
\ee
which implies that $g_0$ is independent of $s$, as before.

To shorten our formulae we will introduce an abbreviated notation. The gauge
potentials $\BA$ and $\BA'$ each carry the set of indices $A,\alpha, \xx$ corresponding to 
the gauge group, spatial rotations and position. These will be contracted with 
kernels which carry several sets of such indices. We will denote such a contraction over a
single pair of sets of indices, followed by
symmetrisation of the result under interchange of the remaining such sets of indices
by a dot. Thus we expand $\cal B$ as
\be
{\cal B}={\cal B}_2+{\cal B}_3+{\cal B}_4=
\BA\cdot(\BA\cdot V_2)+\BA\cdot(\BA\cdot(\BA\cdot V_3))
+\BA\cdot(\BA\cdot(\BA\cdot(\BA\cdot V_4)))
\ee
and $W_0[{\bf A},\BA']$ as

\bq
W_0[{\bf A},\BA']&=&\BA\cdot(\BA\cdot \Xi_2)+\BA\cdot(\BA\cdot(\BA\cdot \Xi_3))
+\BA\cdot(\BA\cdot(\BA\cdot(\BA\cdot \Xi_4)))+..\nn
&&\BA'\cdot(\BA'\cdot \Xi_2)+\BA'\cdot(\BA'\cdot(\BA'\cdot \Xi_3))
+\BA'\cdot(\BA'\cdot(\BA'\cdot(\BA'\cdot \Xi_4)))+..\nn
&&+\BA'\cdot(\BA\cdot \Upsilon_2)+..
\eq
Substituting this into the \Sc equation and extracting the term 
quadratic in $\BA$ gives
\be
\lim_{\lambda\rightarrow\infty}R'_\lambda
\left(
{1\over g_0^2}\BA^\rho\cdot(\BA^\rho\cdot \dot\Xi_2) -2(\BA^\rho\cdot\Xi_2)\cdot(\BA^\rho\cdot\Xi_2)           +{1\over 2g_0^4}\BA^\rho\cdot(\BA^\rho\cdot V_2)\right)=0
\label{sf2}
\ee
where $\dot\Xi$ denotes the derivative of $\Xi$ with respect to $\tau$.
We now assume an expansion of $\Xi_2$ of the form
\be
\Xi_2={1\over g_0^2\tau}\sum_{n=0}^\infty b_n (\tau^2 V_2\cdot)^n 1
\label{xi2}
\ee
which is an expansion in powers of derivatives, since $V_2$ is a 
differential operator. $1$ is the identity kernel. The powers of $\tau$ are fixed by dimensions, 
and the factor of $1/g^2$ is for convenience. Using (\ref{xi2}) in
(\ref{sf2}) yields
\be
\lim_{\lambda\rightarrow\infty}R'_\lambda
\rho\,\left(
{1\over \tau^2}\sum_{n=0}^\infty b_n (2n-1)({\tau^2\over\rho} V_2\cdot)^n 
-{2\over \tau^2}\sum_{n,m=0}^\infty b_n \,b_m\,({\tau^2\over\rho} V_2\cdot)^{n+m} 
+{1\over 2\rho}V_2\cdot \right)\, 1=0
\ee
By equating to zero the coefficients of the various powers of 
$V_2\cdot$ we obtain the $b_n$ as
\be
b_0=-{1\over 2},\quad b_1=-{1\over 6},\quad b_n={2\over 2n+1}\sum_{p=1}^{n-1}b_{n-p}\,b_p,\quad
n>1\label{bee}
\ee
Although our method does not require it, 
in this simple case it is easy to sum the series for $\Xi_2$ to 
give
\be
\Xi_2={ 1\over 2g_0^2}{1+e^{2\tau \omega\cdot}\over 1-e^{2\tau \omega\cdot }} \omega
\label{sumxi}\ee
where $\omega^2=V_2$.
So we can see clearly that as $\tau\rightarrow\infty$
this becomes $-\omega/(2g_0^2)$ which agrees with the result for the vacuum functional
given in (\ref{G2}).

We will now use this to illustrate the reconstruction of the \Sc functional from
its local expansion.  That part of the tree-level functional that we have just obtained
takes the following form when written in terms of Fourier transforms
\bq && 
\BA\cdot(\BA\cdot \Xi_2)\nn
&&={1\over g^2_0\tau}\int {d^3p\over(2\pi)^3}A_\alpha^A (\pp )\,  A_\beta^A (- \pp ) \,\,\left(b_0\,\delta^{\alpha\beta}+\sum_{n=1}^\infty b_{n}\,  ( \del^{\alpha \beta} -
p^\alpha p^\beta/\pp ^2)\,(\tau^2\,
\pp^2)^n \right)
\eq
From (\ref{sumxi}) we see that the infinite series converges
only for $|\pp|\tau<\pi$, but from this series we can reconstruct 
the functional for arbitrary $\pp$ using (\ref{eeeee}). We first write the 
functional for scaled sources after a change of integration variable as

\bq && 
\sqrt\rho\,\BA^\rho\cdot(\BA^\rho\cdot \Xi_2))\nn
&&={1\over g^2\tau}\int {d^3p\over(2\pi)^3}A_\alpha^A (\pp )\,  A_\beta^A (- \pp ) \,\rho\,(b_0\,\delta^{\alpha\beta}+\sum_{n=1}^\infty b_{n}\, (\del^{\alpha \beta} -
p^\alpha p^\beta/\pp ^2 )\,({\tau^2\,
\pp^2\over \rho})^n )
\eq
To do the integration in (\ref{I}) we write $\rho$ as $(\rho-1)+1$
and expand in inverse powers of $(\rho-1)$, resulting in expansion coefficients $\tilde b_n$ that are functions of the product $\tau|\pp|$, thus
\bq
&&\rho\left(
b_0\,\delta^{\alpha\beta}+\sum_{n=1}^\infty b_{n}\, ({\tau^2\,
\pp^2\over \rho})^n  (\del^{\alpha \beta} -
p^\alpha p^\beta/\pp ^2 )\right)\nn
&&
=(b_0(\rho-1)+b_0)\delta^{\alpha\beta}/\tau+\sum_{n=0}^\infty {{\tilde b}_n(\tau|\pp|)\over (\rho-1)^{n}}( \del^{\alpha \beta} -
p^\alpha p^\beta/\pp ^2)
\eq
Using this in (\ref{eeeee}) yields
\bq && 
\BA\cdot(\BA\cdot \Xi_2)\nn
&&=\lim_{\lambda\rightarrow\infty}{1\over g^2\tau}\int {d^3p\over(2\pi)^3}A_\alpha^A (\pp )\,  A_\beta^A (- \pp ) \times\nn
&&\times\left(\quad b_0\delta^{\alpha\beta}+\sum_{n=0}^\infty {\lambda^{n}{\tilde b}_n(\tau|\pp|)\over \Gamma(n+1)\Gamma(n/2+1)}(\del^{\alpha \beta} -p^\alpha p^\beta/\pp ^2 )\right)
\eq
A good approximation to the limit of the infinite series is obtained by truncating the series and taking $\lambda$ as a large as is consistent with the truncation. Consider the finite sum

\be
S_N(\lambda,\tau)={1\over\tau}\left(b_0+\sum_{n=0}^{N-1} {\lambda^{n}{\tilde b}_n(\tau)\over \Gamma(n+1)\Gamma(n/2+1)}\right).
\ee
This is readily computed using the relations (\ref{bee}).
We expect this to provide a good approximation to $S_\infty(\lambda,\tau)$
for values of $\lambda$ for which the last term makes a small contribution to 
the whole.

\begin{figure}[h]
\unitlength1cm  
\begin{picture}(14,10)
\put(6.25,6.35){$\tau=3$}
\put(6.25,5.5){$\tau=2$}
\put(6.25,3){$\tau=1$}
\put(9,0.25){$\lambda$}
\put(5.1,7.5){$S_{N}$}

\centerline{\epsfysize=12cm
\epsffile{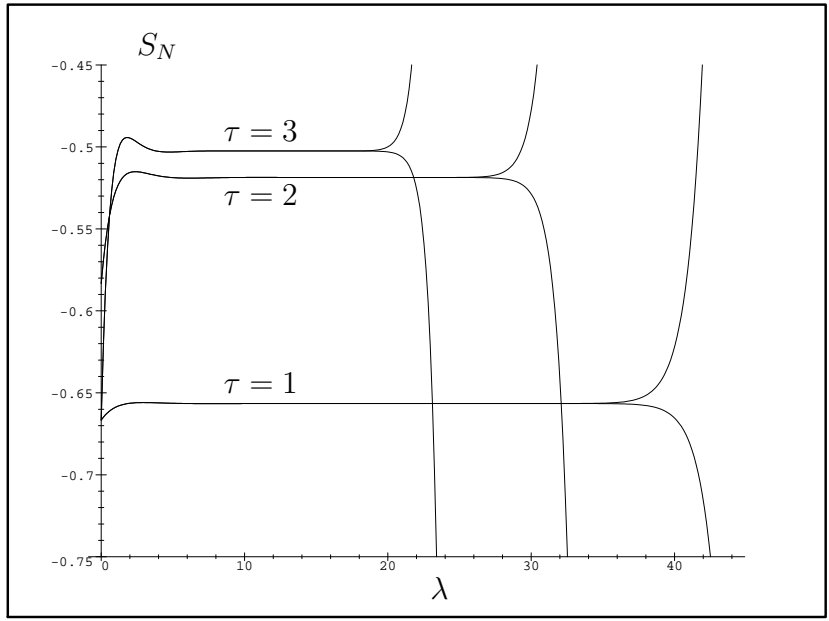} }
\end{picture}

\caption[wave1]{\label{wave1} The truncated series as a function of $\lambda$.}
\end{figure}

In Figure 1 we have plotted $S_{40}(\lambda,\tau)$ and $S_{39}(\lambda,\tau)$ for three values of $\tau$, namely $\tau=1,2,3$.
The last term in $S_{40}(\lambda,\tau)$ is significant where each pair of 
curves can be seen to diverge. The flatness of the curves for slightly smaller values
of $\lambda$ shows that there the value of the truncated series 
provides a good approximation to the value of the full series as
$\lambda\rightarrow\infty$. We have used this to estimate the 
limit in Figure 2 where we have plotted $S_{40}(\tau,\lambda)$ with 
$\lambda$ chosen so that the magnitude of the last term in this series is 
equal to $1/200$, which represents about one per cent of the 
value of the series for large $\tau$, (varying this fraction by an order of magnitude either way makes little difference). Now from (\ref{sumxi})
we should have that the limit as $\lambda\rightarrow\infty$ of 
$S_\infty(\lambda,\tau)$ be $(1+e^{2\tau})/(2(1-e^{2\tau}))\equiv f(\tau)$
and this curve is also shown in Figure 2. We can see that $S_{40}$
provides a good approximation for values of $\tau$ up to about 20
which is a long way beyond the region of convergence of the original series
$\sum_0^{39} b_n\tau^{2n-1}\equiv S_{\rm orig}$ from which we have built $S_{40}$
which is also shown in the Figure diverging from the other two curves at $\tau=\pi$.

\begin{figure}[h]
\unitlength1cm  
\begin{picture}(14,10)
\put(8.5,7.5){$\tau$}
\put(12.3,4.4){$f$}
\put(5.7,5.7){$S_{\rm orig}$}
\put(12.3,3.4){$S_{40}$}
\centerline{\epsfysize=12cm
\epsffile{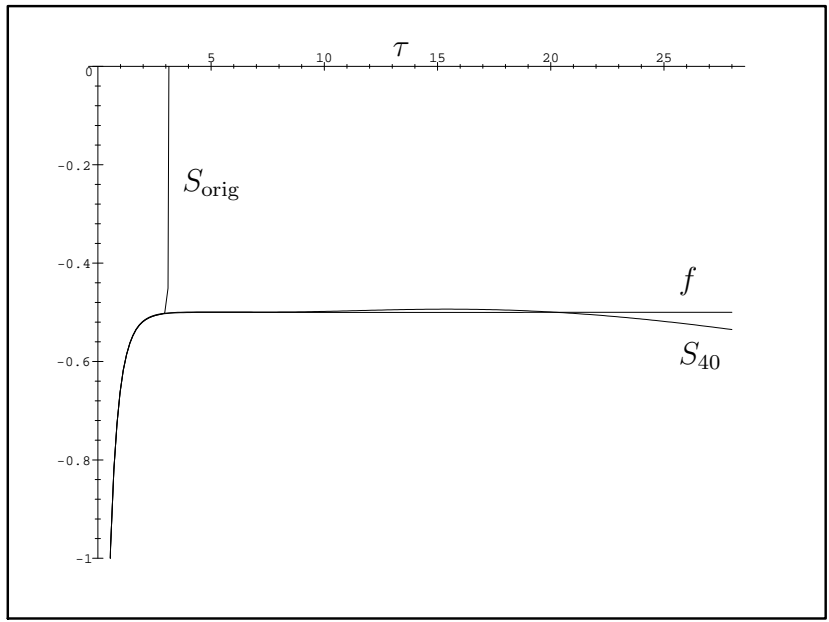} }
\end{picture}
\caption[wave2]{\label{wave2} Tree-level \Sc functional: $\Xi_2$}
\end{figure}

In Appendix C we calculate the coefficients $c_n$ in the local expansion of $\Xi_3$. 
\be
\Xi_3={\tau\over g_0^2}\sum_{n=0}^\infty c_n (\tau^2 V_2\cdot)^n V_3
\label{naxi3}
\ee
and the coefficients $f_n$ and $h_{p,q,r}$ in the local expansion of $\Xi_4$

\be
\Xi_4={\tau\over g_0^2}\sum_{n=0}^\infty f_n (\tau^2 V_2\cdot)^n V_4
+{\tau\over g_0^2}\sum_{p,q,r} h_{p,q,r} \tau^{2(p+q+r+1)} (V_2\cdot)^p
\left(\left((V_2\cdot)^qV_3\right)\cdot\left((V_2\cdot)^rV_3\right)\right) 
\label{naxi4}
\ee
In principle we could use this result to compute the one-loop beta-function.
The combinatorics are rather daunting compared to the approach we will describe in the next section.

\section{Background Field approach, and the beta-function}

Recently Zarembo has shown how to calculate the beta-function for 
Yang-Mills theory using the \Sc equation for the vacuum functional
with a background field method \cite{Zarembo}. As usual the background field approach works especially well at one-loop by reducing the problem to
the evaluation of certain coefficents in the asymptotic expansion of a heat-kernel. We will extend this approach to the \Sc functional
and use it to show that the usual beta-function is obtained from our
local expansion. This is important because the beta-function is sensitive only to high momentum modes and so obtaining the correct result shows that we can indeed reconstruct rapidly varying modes from the slowly 
varying modes of the local expansion using analyticity.
The basic idea is to write $\BA$ and $\BA'$ as fluctuations, $\ba$
and $\ba'$, about a background field ${ \BA_0}$:

\be 
\BA={ \BA_0}+\ba ,\quad \BA'={ \BA_0}+\ba'.
\ee
$\ba$ is now the dynamical variable so

\be
{\delta\over\delta A}={\delta\over\delta a}.
\ee

The potential term in the \Sc equation, $\cal B$ can be expanded 
up to terms quadratic in $\ba$ as
\be
{\cal B}={\cal B}_0+\ba\cdot(\ba\cdot \Lambda)+{\rm Order}(\ba ^3)
\label{calB}
\ee
where ${\cal B}_0$ is the integral of the square of the magnetic field made from $\BA_0$, i.e. ${\bf B}_0=\nabla\wedge \BA_0+\BA_0\wedge \BA_0$,
and
\bq
\Lambda\ba&=&
-\DC\cdot\DC\ba +\DC\DC\cdot\ba
+2{\bf B}_0\wedge\ba+2\ba\wedge{\bf B}_0\nn
&\equiv&-\DC\cdot\DC\ba +\DC\DC\cdot\ba
-2{\bf F}\cdot\ba,
\eq
where $\cal D$ is given by ${\cal D}\omega=[\nabla+{\bf A}_0,\omega ]$.
There is no term linear in $\ba$ since we will take $ \BA_0$ to satisfy
the three-dimensional Yang-Mills equation $\DC\wedge \BA_0=0$.
Consequently 
 when we expand $W[\BA,\BA']$ in powers of $\ba$ 
and $\ba'$ there will be no linear terms. If we write the tree-level contribution to 
$W[\BA,\BA']$ that is quadratic in the fluctuations as $1/\hbar $ times

\be
\ba\cdot(\ba\cdot\Xi)+\ba'\cdot(\ba\cdot\Upsilon)+
\ba'\cdot(\ba'\cdot\Xi)
\ee
then $\Xi$ and $\Upsilon$ are functionals of the background
field $\BA_0$ as well as being functions of $\tau$. Substituting this into the \Sc equation
(\ref{SEQSF2}) and extracting the leading term in $\hbar$ that is
quadratic in $\ba$ leads to the same equation as (\ref{sf2})
but with $\Xi_2$ replaced by $\Xi$ and $V_2$ replaced by
$\Lambda$. Consequently $\Xi$ has the local expansion

\be
\Xi={1\over g_0^2\tau}\sum_{n=0}^\infty b_n (\tau^2 \Lambda\cdot)^n 
\label{xi}
\ee
with the same coefficients, $b_n$, as before. The expansion is valid for fields that vary slowly on the scale of $\tau$.
This tree-level
result gives a one-loop contribution to the renormalisation
of $g^2$ via the condition (\ref{SEQSFR3}). Since $\cal B$ 
has  the expansion (\ref{calB}) it is sufficent
to look at the coefficent of ${\cal B}_0$, and this comes from
$\Delta_s\,(\ba\cdot(\ba\cdot\Xi))$. So to order $\hbar$
\be
{1\over g^4(s)}={1\over g_0^4}+\lim_{\tau\rightarrow\infty}
\hbar \Delta_s\,\left(\ba\cdot(\ba\cdot\Xi)\right)|_{{\cal B}_0}
\ee
If we introduce the mass-scale $\mu$ by setting $s=\rho/\mu^2$,
then the beta-function is
\be
\beta =\lim_{\rho\downarrow 0} \mu{\partial (g^2(\rho/\mu^2))\over\partial \mu}=
\lim_{\tau\rightarrow\infty,\rho\rightarrow 0}
\hbar g^6s{\partial \over\partial s}\Delta_s\,\left(\ba\cdot(\ba\cdot\Xi)\right)|_{{\cal B}_0}
\ee
Direct substitution of (\ref{xi}) into this would be invalid
because as $\rho\rightarrow 0$ the Laplacian includes differentiation
with respect to rapidly fluctuating fields which are absent from the local
expansion. As we have shown, however,
the former can be obtained from the latter by analyticity, and this is encoded into
the $R_\lambda'$ operator. For large $s$ 

\be
\Delta_s\,\ba\cdot(\ba\cdot\Xi)=
2 \, Tr K\,\left({1\over g^2_0\tau}\sum_{n=0}^\infty b_n (\tau^2 \Lambda\cdot)^n\right) ,\label{delxi}
\ee
where $Tr$ denotes the functional trace. Previously we chose
the kernel, $K$, in the Laplacian to have a simple form in terms of a momentum
cut-off multiplied by  a Wilson line, but we are free to make other choices provided that we preserve gauge invariance, that the kernel represents a 
$\delta$-function as the cut-off is removed, and that it scales covariantly. It is particularly convenient for the present calculation to take $K$ to be the heat-kernel of the operator
$\Lambda$ at time $s$, thus $K=\exp (-s\Lambda)$ and so the various powers of
$\Lambda$ acting on the kernel are obtained by differention with respect 
to $s$

\be
s{\partial \over\partial s}\Delta_s\,\ba\cdot(\ba\cdot\Xi)=
2s{1\over g^2_0\tau}\sum_{n=0}^\infty b_n \left(-\tau^2 {\partial\over\partial s}\right)^n {\partial \over\partial s}\,Tr \,\left(K\right) .\label{delxii}
\ee
In Appendix D we use the results of \cite{PYM} 
to show that  
\be
s{\partial \over\partial s}\Delta_s\,\ba\cdot(\ba\cdot\Xi)\Big|_{{\cal B}_0}
={22N\over 3}{1\over (4\pi)^{3/2}}
\sum_{n=0}^\infty b_n {\Gamma (n+1/2)\over \Gamma(1/2)}\left({\tau^2\over s}\right)^{n-1/2}
\ee
If we put $s=\rho/\mu^2$ the series converges only for large $\rho$, but represents
a function that is analytic in the complex $\rho$-plane cut along the negative real axis.
If we multiply by $\sqrt \rho$ the resulting large-$\rho$ series has only integer powers
so the cut for this function has only finite length. Consequently we can obtain the 
small-$s$ value as 
\bq
&&
\lim_{s\downarrow 0}
s{\partial \over\partial s}\Delta_s\,\ba\cdot(\ba\cdot\Xi)\Big|_{{\cal B}_0}
=\nn&&
\lim_{\lambda\rightarrow\infty}
{1\over 2\sqrt\pi i}\int_C {d\rho\over \rho}\,e^{\lambda \rho}
\, \sqrt{\lambda\rho} {22N\over 3}{1\over (4\pi)^{3/2}}
\sum_{n=0}^\infty b_n {\Gamma (n+1/2)\over \Gamma(1/2)}\left({\tau^2\mu^2\over \rho}\right)^{n-1/2}
\eq
If the contour $C$ is collapsed to one around the cut consisting
of a small circle centred on the origin, $C''$, and the rest consisting of two lines slightly above and below the negative real axis, then the contribution of the latter is damped by the 
$\exp(\lambda\rho)$ factor and the contribution of the former yields the left-hand side
because
\be
\lim_{\lambda\rightarrow\infty}
{1\over 2\pi i}\int_{C''} {d\rho\over \rho}\,e^{\lambda \rho}
\, \sqrt{\lambda\rho}=\sqrt\pi\,.
\ee
Performing the contour integration gives
\be
{22N\over 3}{1\over (4\pi)^{3/2}}
\lim_{\lambda\rightarrow\infty}
\sum_{n=0}^\infty b_n {\Gamma (n+1/2)\over \Gamma(n)}\left({\lambda\tau^2\mu^2}\right)^{n-1/2}
\ee
Using the same argument as previously the convergence can be improved by doing a further
contour integral in the complex $1/\sqrt\lambda$ cut along the negative real axis to obtain

\be
{22N\over 3}{1\over (4\pi)^{3/2}}
\lim_{\lambda\rightarrow\infty}{1\over 2\sqrt\pi i}\int_{C''} {d\rho\over \rho}\,e^{\lambda^2 \rho}
\, \sqrt{\lambda^2\rho}
\sum_{n=0}^\infty b_n {\Gamma (n+1/2)\over \Gamma(n)}\left({\tau^2\mu^2\over\sqrt\rho}\right)^{n-1/2}
\ee
which gives the $\beta$-function as

\be
\beta =\hbar g^4 {11N\over 12\pi} 
\lim_{\lambda,\tau\rightarrow\infty}
\sum_{n=0}^\infty b_n {\Gamma (n+1/2)\over\Gamma(n) \,\Gamma(n/2+1/4)}\left({\lambda\tau^2\mu^2}\right)^{n-1/2}.
\ee
\begin{figure}[h]
\unitlength1cm  
\begin{picture}(14,10)

\put(9,1.25){$x$}
\put(12,3){$S_{99}^\beta$}
\put(12,6.5){$S_{100}^\beta$}
\put(7,5.2){$-{1\over 2\pi}$}

\centerline{\epsfysize=12cm
\epsffile{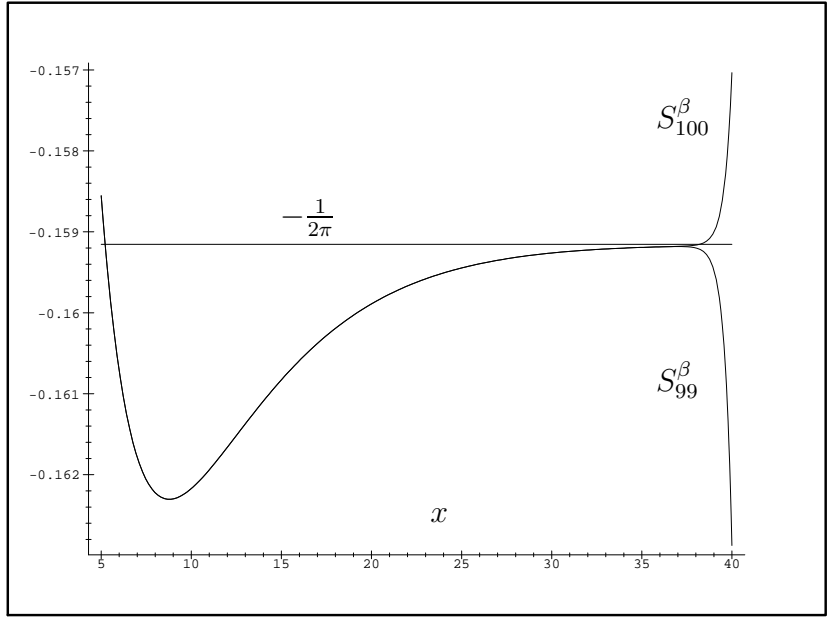} }
\end{picture}
\caption[beta1]{\label{beta1} The series $S_{99}^\beta$ and $S_{100}^\beta$}
\end{figure}
\eject
In Figure 3 we have shown as a function of $x$ the finite sum

\be
S^\beta_N(x)\equiv
\sum_{n=0}^N b_n {\Gamma (n+1/2)\over\Gamma(n) \,\Gamma(n/2+1/4)}x^{n-1/2}.
\ee
for $N=100$ and $N=99$. These sums are seen to diverge for $x\approx 39$ 
which is when the last term of the former series begins to have a
non-negligible value. We can trust the finite sum $S_{100}^\beta$
as a good approximation to $S_\infty^\beta (x)$ up to this value of $x$.
From the known value of the $\beta$-function the limit of this
series should be

\be
\lim_{x\rightarrow \infty} S_\infty^\beta (x)=-{1\over 2\pi}
\ee
and this value is shown as the stright line in the figure.
From this it is clear that the limit is well approximated by taking the value of the finite sums just before they diverge. In Figure 4 we have shown the 
same curves over a shorter range of $x$, from which we can see
that this procedure for estimating the limit gives the numerical value
$S^\beta_{100}(37)=-0.15918$ which as an estimate of $-1/(2\pi)\approx
-0.159155$ is correct to about two parts in $10^{4}$. 
\begin{figure}[h]
\unitlength1cm  
\begin{picture}(14,10)

\put(9,1.25){$x$}
\put(12,3){$S_{99}^\beta$}
\put(11.7,6){$S_{100}^\beta$}
\put(7,5.5){$-{1\over 2\pi}$}

\centerline{\epsfysize=12cm
\epsffile{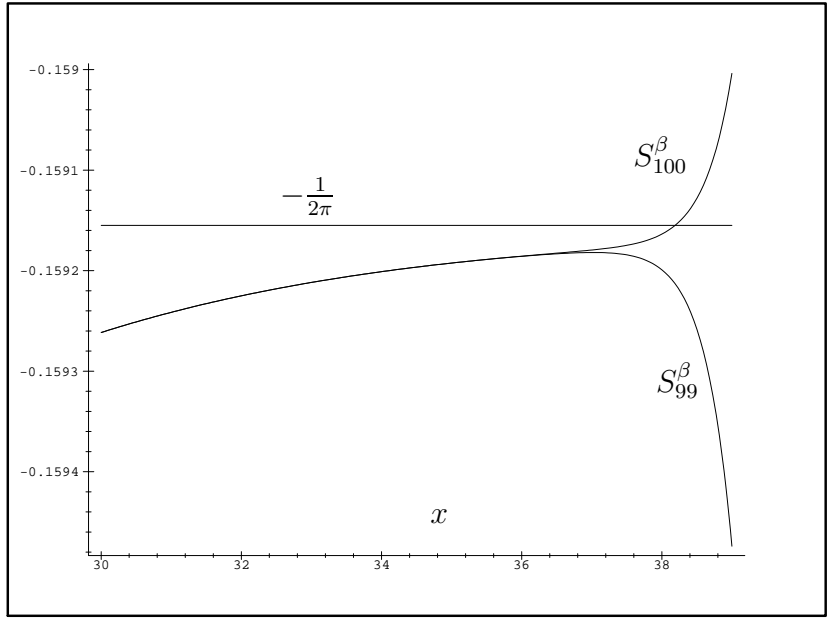} }
\end{picture}
\caption[beta2]{\label{beta2} The series $S_{99}^\beta$ and $S_{100}^\beta$}
\end{figure}
The error may be estimated {\it a priori} by examining how the curve $S^\beta_{100}$
approaches the straight line that represents its limiting value.
From the method we employed to suppress the contribution of the cut 
we would expect that the deviation from a stright-line 
is roughly exponential in $\lambda$ for large $\lambda$.
In Figure 5 we have shown the logarithm of the derivative of $S^\beta_{100}(x)$
which shows that this is indeed approximately linear in $x$ with slope $\approx -1/5$,
hence $S^\beta_{100}\approx A+B\exp(-x/5)$. The error in approximating 
$\lim{\lambda\rightarrow\infty}\quad S_\infty^\beta$ by $S^\beta_{100}(37)$
is thus of the order of $B\exp(-x/5)\approx |5\,S^{\beta '}_{100}(37)|\approx
2.4\times 10^{-4}$.

\begin{figure}[h]
\unitlength1cm  
\begin{picture}(14,10)

\put(9,1.25){$x$}
\put(9.5,4.6){$\log (S_{100}^{\beta'}(x))$}

\centerline{\epsfysize=12cm
\epsffile{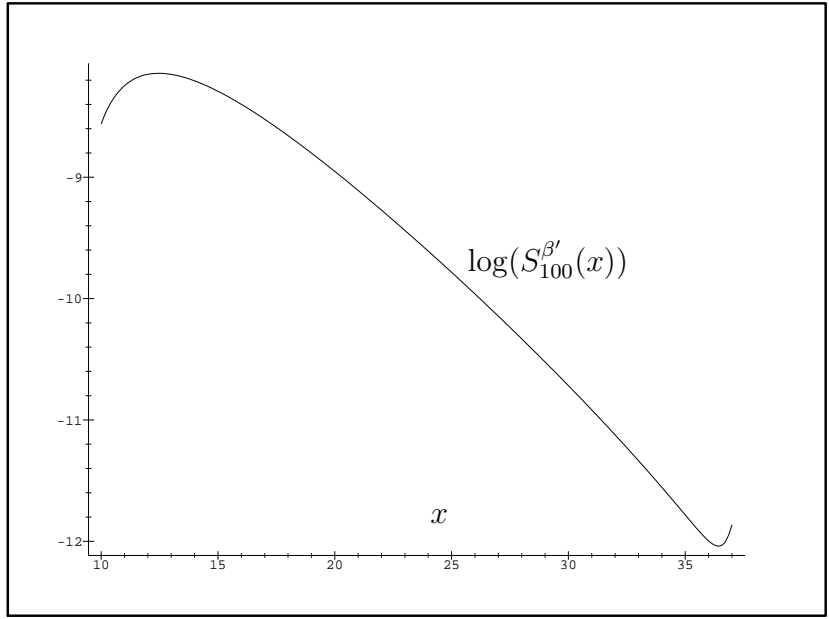} }
\end{picture}
\caption[errr]{\label{errr} $\log (S_{100}^{\beta'}(x))$}
\end{figure}

\section{Conclusions}
When a scaling ${\bf x}\rightarrow {\bf x}/\sqrt \rho$ is applied to a quantum field theory many quantities can be continued to analytic
functions in the complex $\rho$-plane cut along the negative real axis.
We have seen this to be the case for the connected Green's functions that
contribute to the logarithm of the \Sc functional, and also for appropriately defined bare couplings, (and consequently their beta-functions). It is
also a property of, for example, the two-point function of Lorentz invariant operators as
can be seen from the K\"allen-Lehmann representation, and of the Zamolodchikov c-function. We have exploited this 
analyticity to reconstruct the \Sc functional for arbitrarily varying fields
from its expression as a local expansion that is valid for slowly varying fields, and we have obtained the form of the \Sc equation that acts directly on
this local expansion. This leads to coupled {\it algebraic} equations that may,
in principle, be solved without recourse to perturbation theory, but our
purpose here has been to show that when these equations are solved in powers of
the coupling they are in agreement with the expected results of perturbation theory. To make this comparison we first solved the functional \Sc equation
in a `standard' perturbative manner without using the local expansion.
We also computed the beta-function in both approaches and obtained the usual result. This illustrates that short-distance effects can be correctly reproduced
from our small momentum expansion because the beta-function is a consequence of
rapidly varying field configurations.


\appendix
\section{Tree-level contribution to $W[\BA]$}
In this appendix we 
calculate the tree-level contributions to $W[\BA]$ up to fourth order,
i.e. $W_0^2$, $W_2^3$ and $W_0^4$.

\subsection{Calculating $W_0^2$}

The Hamilton-Jacobi equation gives 
\be
g_0^4\int {d^3p\over (2\pi)^3} \frac{\del W_0^2}{\del A^R_\alpha (\pp)}
\frac{\del W_0^2}{\del A^R_\alpha (- \pp)} =  {\cal
B}_2  \, ,
\label{EQN:B2}
\ee
in which ${\cal B}_2$ is the part of the potential ${\cal B}$
that is quadratic in $A$:
\bq
{\cal B}_2 &=& \int {d^3p\over(2\pi)^3} (\pp \wedge \BA ^A (\pp )) \cdot (\pp \wedge \BA ^A (- \pp ))
\nonumber \\
&=&  \int {d^3p\over(2\pi)^3} A_\alpha^A (\pp )  A_\beta^A (- \pp ) \,\, (\pp ^2 \del^{\alpha \beta} -
p^\alpha p^\beta) \,\, ,
\eq
We can take $W_0^2$ to have the form
\be
W_0^2 = {1\over g_0^2}\int {d^3p\over(2\pi)^3} A_\mu^A(\pp ) A_\nu^A( - \pp ) \,\, \Gamma^{\mu \nu} (\pp ),
\quad \Gamma^{\nu \mu } (-\pp )=\Gamma^{\mu \nu} (\pp )
\ee
Hence
\be
 \frac{\del W_0^2}{\del A^R_\alpha (\pp)} =  {2\over g_0^2}\Gamma^{\alpha \nu} (\pp) A_{\nu}^R
(-\pp)
\label{DW2}
\ee
so that the Hamilton-Jacobi equation and Gauss' law become

\be 4 \Gamma^{\mu \alpha} (\pp ) \delta_{\alpha\beta} \Gamma^{\beta \nu} (\pp )=
\pp ^2 \del^{\mu\nu} -
p^\mu p^\nu\,,\quad p_\mu\Gamma^{\mu \nu} (\pp )=0
\ee
which has the solution
\be
\Gamma_{\alpha \beta} ( \pp ) = 
\frac{p_\alpha p_\beta - \pp^2 \del_{\alpha \beta}}{2 |\pp |}.
\label{G2}
\ee
This is unique up to a sign which is chosen to make $\Psi$ square integrable to this order.


\subsection{Calculating $W_0^3$} 

To third order in $\BA$ the Hamilton-Jacobi equation is
\be
2 g_0^4\int {d^3p\over(2\pi)^3} \frac{\del W_0^2}{\del A^R_\eta (- \pp)}
 \frac{\del W_0^3}{\del A^R_\eta ( \pp)} ={\cal B}_3 
\label{SEW3}
\ee
with
\be
 {\cal B}_3 = - i  \, \int {d^3p_1\,d^3p_2\,d^3p_3\over(2\pi)^6}
 \del^3 (\pp_1 + \pp_2 + \pp_3) (\pp_1 \wedge \BA^A(\pp_1) )\cdot
(\BA^B(\pp_2) \wedge \BA^C (\pp_3)) \,\, f^{ABC}  \, .
\ee
$W_0^3$ has the form

\be
W_0^3 = {1\over g_0^2}\int {d^3p_1\,d^3p_2\,d^3p_3\over(2\pi)^6}
A_\alpha^A(\pp_1)A_\beta^B(\pp_2)A_\gamma^C(\pp_3)\,\,\, f^{ABC} \, \Gamma^{\alpha
\beta \gamma } (\pp_1,\pp_2,\pp_3)\,  \del^3 (\pp_1 + \pp_2 + \pp_3) \,\, .
\ee
The sign of $\Gamma^{\alpha
\beta \gamma } (\pp_1,\pp_2,\pp_3)$ changes if we simultaneously 
interchange any two momenta and their respective indices, as it does if we reverse all the momenta. (\ref{SEW3}) becomes 
\bq
&& \int{d^3p_1\,d^3p_2\,d^3p_3}\,
A_\alpha^A(\pp_1)A_\beta^B(\pp_2)A_\gamma^C(\pp_3) \,\,  \,\,  \del^3 (\pp_1 + \pp_2 + \pp_3) \nonumber \\
&& {\Big\{ }f^{ABC}\left(
 12 \Gamma_\eta ^\alpha (\pp_1) \Gamma^{\eta \beta \gamma} (\pp_1, \pp_2,
\pp_3)  +  i  p_{1  \eta} \delta^{\eta \alpha}_{\beta \gamma} \right){\Big\} } = 0 \, ,
\eq
implying
\be
{\cal S}{\Big\{ }f^{ABC}\left(
 12 \, \Gamma_{\eta \alpha} (\pp_1)
 \Gamma_{\eta \beta \gamma} (\pp_1, \pp_2, \pp_3)  + 
 i  p_{1  \eta} \delta^{\eta \alpha}_{\beta \gamma} \right){\Big\} } = 0 \, . 
\label{SEW3-SYM}
\ee
${\cal S}$ standing for symmetrisation under interchange of the sets $\{ A,\alpha,\pp_1\}$,
 $\{ B,\beta,\pp_2\}$,  and $\{ C,\gamma,\pp_3\}$, giving
\be
4 f^{ABC}{\Big\{} \Gamma_{\alpha \eta} (\pp_1) 
\Gamma_{\eta \beta \gamma} (\pp_1, \pp_2, \pp_3) +  \Gamma_{\beta \eta} (\pp_2) 
\Gamma_{\alpha \eta \gamma} (\pp_1, \pp_2, \pp_3) +  \Gamma_{\gamma \eta} (\pp_3) 
\Gamma_{\alpha \beta \eta} (\pp_1, \pp_2, \pp_3)  {\Big\}}
\ee 

 for the first term between braces and
\be
\frac{1}{3}  f^{ABC}{\Big\{} p_{1  \eta} \delta^{\eta \alpha}_{\beta \gamma} + 
p_{2  \eta} \delta^{\eta \beta}_{\gamma \alpha} +
p_{3  \eta} \delta^{\eta \gamma}_{\alpha \beta}  {\Big\}}
\ee
for the second.
Gauss's law  (\ref{GL})  for $W_0^3$ states that
\be
-i \pp . \frac{\del W_0^3}{\del \BA^R (-\pp)} = -  f^{RST} \int{d^3q\over(2\pi)^3} \BA^S (\pp - \qq)
  \frac{\del W_0^2}{\del \BA^T (-\qq)} 
\ee 

which can be written as
\be
p_{1 \alpha} \Gamma_{\alpha \beta \gamma} (\pp_1, \pp_2, -\pp_1 - \pp_2 ) = \frac{i}{3} \Big(
\Gamma_{\beta \gamma} (\pp_3) - \Gamma_{\beta \gamma}(\pp_2) \Big) \,\,\,
\label{GL3}
\ee
(\ref{GL3}) and (\ref{SEW3-SYM}) have the solution
\bq
&&\Gamma_{\alpha \beta \gamma}(\pp_1, \pp_2, \pp_3) = 
\nonumber \\&&
\frac{i }{3
(|\pp_1|+|\pp_2|+|\pp_3|)} \times {\Bigg\{} + \hat{p}_{1 \alpha} (\Gamma_{\beta \gamma}(\pp_3)  -  \Gamma_{\beta \gamma}(\pp_2) )
-\hat{p}_{2 \beta} (\Gamma_{\alpha \gamma}(\pp_3)  \nonumber \\
&-&   \Gamma_{\alpha \gamma}(\pp_1) )   
- \hat{p}_{3 \gamma} (\Gamma_{\alpha \beta}(\pp_1)  -  \Gamma_{\alpha \beta}(\pp_2) ) 
+ \frac{1}{2} ( p_{1 \eta} \delta^{\eta \alpha}_{\beta \gamma} +
                  p_{2 \eta} \delta^{\eta \beta}_{\gamma \alpha} +
                  p_{3 \eta} \delta^{\eta \gamma}_{\alpha \beta} )  
 {\Bigg\}}
\label{G32}
\eq  


\subsection{ Calculating $W_0^4$ }
The quartic terms in the Hamilton-Jacobi equation yield:
\be
g_0^4\int {d^3p\over(2\pi)^3} \Bigg( 2 \frac{\del W_0^2}{\del A^R_\alpha (- \pp)}
                    \frac{\del W_0^4}{\del A^R_\alpha ( \pp)}
                 +  \frac{\del W_0^3}{\del A^R_\alpha (- \pp)}
                     \frac{\del W_0^3}{\del A^R_\alpha ( \pp)} \Bigg)
                 = {\cal{B}}_4  .
\label{SEW4}
\ee
We take $W_0^4$ to have the form
\bq
W_0^4 &=& {1\over g_0^2}\int{d^3p_1\,d^3p_2\,d^3p_3\,d^3p_4\over(2\pi)^9} \del^3 ( \pp_1 + \pp_2 + \pp_3 +
\pp_4 ) \, A_\alpha^A (\pp_1) A_\beta^C (\pp_2) A_\gamma^B (\pp_3) A_\delta^D (\pp_4)
\times \nonumber \\
&\times& f^{MAB}f^{MCD}\Gamma_{\alpha \beta \gamma \delta} (\pp_1, \pp_2, \pp_3, \pp_4) \, .
\label{W4}
\eq
The quartic term of the potential is
\bq
{\cal{B}}_4 &=&   
 {1\over 4}\int{d^3p_1\,d^3p_2\,d^3p_3\,d^3p_4\over(2\pi)^9} \del^3 ( \pp_1 + \pp_2 + \pp_3 +
\pp_4 ) \, A_\alpha^A (\pp_1) A_\beta^C (\pp_2) A_\gamma^B (\pp_3) A_\delta^D (\pp_4)
\times \nonumber \\
&\times& f^{MAB}f^{MCD} \del^{\alpha \beta}_{\gamma \delta} \, .
\eq
The Hamilton-Jacobi equation implies that
\bq
&&{\cal{S}}_4{\Big\{} f^{MAB}f^{MCD} \Big(16 \Gamma_{\alpha \eta}(\pp_1)\Gamma_{\eta \beta \gamma \delta}
 (\pp_1, \pp_2, \pp_3, \pp_4) +\nonumber \\ && 
 9 \, \Gamma_{\eta \alpha
\beta}(\pp_3 + \pp_4, \pp_1, \pp_2)\Gamma_{\eta \gamma \delta}(\pp_1 + \pp_2, \pp_3,
\pp_4)-  {1\over 4}\del^{\alpha \beta}_{\gamma \delta} \Big){\Big\}}  =  0 \, ,
\label{SEW4-S}
\eq 
Where ${\cal S}_4$ symmetrises the expression under simultaneous interchanges of the four sets 
of indices and momenta $A,\alpha,\pp_1$ etc.
Gauss's law for $W_0^4$,
\be
i \pp . \frac{\del W_0^4}{\del \BA^R ( \pp)} = -  f^{RST} \int {d^3q\over (2\pi)^3} \BA^S (-\pp - \qq)
  \frac{\del W_0^3}{\del \BA^T (-\qq)} \, ,
\ee
can be expressed as
$$
{\cal S}_3\left\{f^{MNA}f^{MCD}\left(4p_\eta \Gamma_{\eta \alpha \beta \gamma} (\pp, \pp_1, \pp_2, \pp_3)  - 3
i \,\, \Gamma_{\alpha \beta \gamma}(\pp+\pp_1, \pp_2 , \pp_3)\right) \right\}
$$
\be
{\mbox{when}} \,\,\,\,\,\, \pp + \pp_1 + \pp_2 + \pp_3 = 0 \, ,
\label{GLG4}
\ee
and ${\cal S}_3$ denotes symmetrisation under interchange of the three sets  $\{A,\alpha,\pp_1\}$,
$\{B,\beta,\pp_2\}$ and $\{C,\gamma,\pp_3\}$. 
Using this in (\ref{SEW4-S}) we find

\bq
&&{\cal{S}}_4{\Big\{} f^{MAB}f^{MCD} \Gamma_{\alpha \beta \gamma \delta}
 (\pp_1, \pp_2, \pp_3, \pp_4)\Big\} =\nonumber \\ && 
{\cal{S}}_4{\Big\{} f^{MAB}f^{MCD} \omega
\Big( 6i{\hat{p}}_{1 \alpha}  \Gamma_{\beta \gamma \delta}(\pp_1 +\pp_2,  \pp_3, \pp_4) 
 +\nonumber \\ && \,\,\,\,\,\,\,\,\,\,\,\
+9 \, \Gamma_{\eta \alpha
\beta}(\pp_3 + \pp_4, \pp_1, \pp_2)\Gamma_{\eta \gamma \delta}(\pp_1 + \pp_2, \pp_3,
\pp_4)-  {1\over 4}\del^{\alpha \beta}_{\gamma \delta} \Big){\Big\}}  
\label{G42}
\eq
where 
\be
\omega={1\over 2(|\pp_1| + |\pp_2| + |\pp_3| + |\pp_4|)}
\ee


\section{Contributions to the one loop Beta-function}
In this appendix we calculate the small $s$ behaviour of $\Delta_s W_0^2$, $\Delta_s W_0^3$ and 
$\Delta_s W_0^4$.


\subsection{$\Delta_s W_0^2$}
\label{SS-LW2}

Since $W_0^2$ is only quadratic in $\BA$ the $\BA$-dependence of $\Delta_s W_0^2$
derives from the Wilson line in the reulator kernel.
The term that is quadratic in $A$ is thus
\be
\Big( \Delta_s g_0^2\,W_0^2 \Big)_{A^2} = \int_{|\pp|^2<1/s} {d^3x\,d^3y\,d^3q\,d^3p\over (2\pi)^6} e^{i\, \pp (\xx
- \yy)} \Big[ U^{RS}(\xx, \yy)\Big]_{A^2} \, 2 \, e^{- i\, {\bf{q}}  (\xx - \yy)}
\del^{RS} \Gamma_{\rho \rho} ( q ) \, ,
\ee
where 
\be
\Big[ U^{RS}(\xx, \yy)\Big]_{A^2} = {\cal{P}} \, \Bigg( 
\frac{1}{2} \int_{\xx}^{\yy}d{\mbox{w}}_{\mu} \int_{\xx}^{\yy}
d{\mbox{z}}_{\nu} \, A^A_{\mu}({\bf{w}})
 A^B_{\nu}({\bf{z}})  \, T^A T^B  \Bigg)_{RS} \, .
\ee
We parametrise the integration path as
\bq
{\bf{w}} &=& \xx + \sigma_1 ( \yy - \xx ) \, , \quad 0 \le \sigma_1 \le 1 \nonumber \\  
{\bf{z}} &=& \xx + \sigma_2 ( \yy - \xx ) \, , \quad 0 \le \sigma_2 \le 1 \, ,
\eq
and change the variables such that $\yy \rightarrow 
{\bf{v}} = ( \yy - \xx )$. Then, after integrating over $\xx$, we get
\bq
&&
\Big( \Delta_s g_0^2\,W_0^2 \Big)_{A^2} =\nonumber\\
&& -  \, \int d^3v\,d^3q\,d^3k\,d^3k'\,d^3p\,d\sigma_1\,d\sigma_2\,(2\pi)^{-9} \del^3(  {\bf{k}} 
+  {\bf{k}}' )
 \,  A_\mu^A({\bf{k}}) A_{\nu}^B({\bf{k}}') \nonumber\\
&& \times e^{i\, {\bf{v}}. (\qq + \sigma_1 {\bf{k}} + \sigma_2 {\bf{k}}' - \pp )} 
|\qq | \, {\mbox{v}}_\mu  {\mbox{v}}_\nu (T^A T^B)_{RR} \, ,
\label{LW2}
\eq
where the integration is restricted by  $
|\pp|^2 < 1/s, 0 \le \sigma_1 \le 1,  0 \le \sigma_2  \le 1 $. Finally, integrating
(\ref{LW2}) over ${\bf{k}}'$ and ${\bf{v}}$ allows us to write \footnote{ We use
 the identities $f^{RAC}f^{CBR} = - N \, \del^{AB}$ and 
${\mbox{v}}_\mu  {\mbox{v}}_\nu \, exp (i {\bf{v}}. \qq) = 
- \frac{\der^2}{\der q_\mu \der q_\nu} \, exp (i {\bf{v}}. \qq)$ \, .}
\bq &&
\Big( \Delta_s g_0^2\,W_0^2 \Big)_{A^2} = - {N\over  (2 \pi)^6} \, 
\int \int d^3q\,d^3k\,d^3p\,d\sigma_1\,d\sigma_2\, A_\mu^A({\bf{k}}) A_\nu^A (-{\bf{k}}) 
\nonumber\\
&& 
\frac{\der ^2}{\der q_\mu \der q_\nu} \Bigg[ \del ^3 
( \qq + {\bf{k}}(\sigma_1 - \sigma_2) -\pp ) \Bigg] \,\, 
|\qq |  \, ,
\eq  
We can integrate by parts to make the two derivatives
act on $|\qq|$, and use 
\be
\frac{\der^2 |\qq|}{\der q_\mu \der q_\nu} = \frac{1}{|\qq|} \Bigg( \del_{\mu \nu}
- \frac{q_\mu q_\nu}{|\qq|^2} \Bigg).
\ee

This gives
\be
- {N \over (2 \pi)^6} \, 
\int d^3q\,d^3k\,d^3p\,d\sigma_1\,d\sigma_2\, A_\mu^A({\bf{k}}) A_\nu^A (-{\bf{k}}) \,\,  \del ^3 
( \qq + {\bf{k}}(\sigma_1 - \sigma_2) -\pp ) \,\, \frac{1}{|\qq|} \Bigg( \del_{\mu \nu}
- \frac{q_\mu q_\nu}{|\qq|^2} \Bigg).
\ee
If we perform the integration over $\qq$ using the delta function, call
$\sigma_1 - \sigma_2 = y$ and make a change of variables such that
\be
\int_0^1 d\sigma_2 \int_0^1 d\sigma_1 =  \int_0^1 d\sigma_2 
\int_{-\sigma_2}^{1-\sigma_2} d y
\ee
we end up with
\bq
- N  \, 
\int {d^3k\,d^3p\,d\sigma_2\,dy\over (2\pi)^6}
A_\mu^A({\bf{k}}) A_\nu^A (-{\bf{k}}) &\times& \,\, 
\Bigg\{ \frac{\del_{\mu \nu}}{|\pp - {\bf{k}} y|} - \frac{p_\mu p_\nu}{|\pp - {\bf{k}} 
y|^3} +  \frac{p_\mu k_\nu y}{|\pp - {\bf{k}} y|^3} \nonumber\\ &+&
\frac{p_\nu k_\mu y}{|\pp - {\bf{k}} y|^3} -  
\frac{k_\mu k_\nu y^2}{|\pp - {\bf{k}} y|^3}
\Bigg\}
\nonumber\\ \equiv
\int d^3k\, A_\mu^A({\bf{k}}) A_\nu^A (-{\bf{k}})\Lambda^{\mu\nu}_2({\bf{k}})&&
\label{C1}
\eq
where $ -\sigma_2 \le y \le 1 -
\sigma_2$, $ 0 \le \sigma_2 \le 1 $ and $|\pp|^2<1/s$.


\subsection{$\Delta_s W_0^3$}
\label{SS-LW3}

In this case, the quadratic term is obtained by taking the term that is linear in $A$
in the expansion of $U^{RS} (\xx, \yy)$. Following similar steps as we presented 
in section (\ref{SS-LW2}) and using that
\be
\frac{\del^2 W_0^3}{\del A_\rho^R (\yy) \del A_\rho^S (\xx)} = {6\over g^2_0} \,
\int_{{\bf{z}}} A_\nu^C({\bf{z}}) f^{SRC} \, \Gamma_{\rho \rho \nu} (\xx, \yy,
{\bf{z}}) \, ,
\ee
we obtain
\bq
\Big( \Delta_s g_0^2\,W_0^3 \Big)_{A^2} = -{6 i \, N \over (2 \pi )^6} && \int_{\Sigma ''}d^3k\,d^3q\,d^3p\,
A_\mu^A(-{\bf{k}}) A_\nu^A ( {\bf{k}}) \,\,
\del^3 \Big( \pp - \qq + (\sigma - 1) {\bf{k}} \Big) \, \times \nonumber \\
&& \times
\frac{\der  \Gamma_{\rho \rho \nu} (\qq, -\qq - {\bf{k}}, {\bf{k}})}{\der q_\mu} 
\nonumber \\
&& 
 \equiv
\int d^3k\, A_\mu^A({\bf{k}}) A_\nu^A (-{\bf{k}})\Lambda^{\mu\nu}_3({\bf{k}})
\label{C2}
\eq
where $|\pp |^2 < 1/s$ and $0 \le \sigma \le 1 $.


\subsection{$\Delta_s W_0^4$}
The term that is quadratic in $A$ receives no contribution from the Wilson line
and so is just
\be
\Big( \Delta_s W_0^4 \Big)_{A^2} = \int_{|\pp|^2 < 1/s} {d^3p\over(2\pi)^3}e^{i \pp \, (\xx - \yy)}
\del^{RS} \frac{\del^2 W_0^4}{\del A^R_\rho (\yy) \del A^S_\rho (\xx )} \, .
\label{LW4}
\ee
We obtain from (\ref{G4})

\bq
&&\Big( \Delta_s g_0^2 W_0^4 \Big)_{A^2} = - N \int_{ {\bf{k}}, |\pp|^2 < 1/s}
 A_\mu^A( {\bf{k}})
 A_\nu^A(- {\bf{k}}) \,\,\,  \frac{1}{(|\pp | + | {\bf{k}} |)} 
{\Bigg\{}  \nonumber \\
&&\quad\quad 18 \, \Gamma_{\eta \rho \mu} (\pp +  {\bf{k}}, -\pp, - {\bf{k}})
\Gamma_{\eta \rho \nu} (\pp +  {\bf{k}}, -\pp, - {\bf{k}})  + 6i  {\hat{k}}_{\mu} \Gamma_{\nu \rho \rho}(-{\bf{k}}, \pp + 
 {\bf{k}}, 
-\pp)\nonumber \\
&&\quad\quad -6i  {\hat{p}}_{\rho} \Gamma_{\mu \rho \nu}(\pp+ {\bf{k}}, -\pp , 
 -{\bf{k}}) 
 + 2 \, \del^{\mu \nu}  {\Bigg\}}
\nonumber \\
&& 
\equiv
\int d^3k\, A_\mu^A({\bf{k}}) A_\nu^A (-{\bf{k}})\Lambda^{\mu\nu}_4({\bf{k}})
\label{C3}
\eq

\section{Cubic and Quartic Terms in the \Sc Functional.}

The term in the \Sc equation that is cubic in $\BA$ is

\bq&&\lim_{\lambda\rightarrow\infty}R'_\lambda
\,\Big(
{1\over g^2}\BA^\rho\cdot(\BA^\rho\cdot(\BA^\rho\cdot \dot\Xi_3)) -6\BA^\rho\cdot((\BA^\rho\cdot\Xi_3))\cdot(\BA^\rho\cdot\Xi_2) \nn
&&\quad\quad          +{1\over 2g^4}\BA^\rho\cdot(\BA^\rho\cdot(\BA^\rho\cdot V_3))\Big)=0
\label{sf3}
\eq
We assume an expansion of $\Xi_3$ of the form
\be
\Xi_3={\tau\over g_0^2}\sum_{n=0}^\infty c_n (\tau^2 V_2\cdot)^n V_3
\label{xi3}
\ee
which is again an expansion in powers of derivatives.  Using this in
\ref{sf3} leads to
\bq&&\lim_{\lambda\rightarrow\infty}R'_\lambda
\,\rho\Big({1\over \tau^2}\sum_{n=0}^\infty c_n (2n+1)({\tau^2\over\rho} V_2\cdot)^nV_3 
\nn
&&\quad\quad-{6}\sum_{n,m=0}^\infty b_n \,c_m\,({\tau^2\over\rho} V_2\cdot)^{n+m}V_3 
+{1\over 2\rho}V_3\Big )=0
\eq
By equating to zero the coefficients of the various powers of 
$V_2\cdot$ we obtain the $c_n$ as
\be
c_0=-{1\over 8},\quad c_n={3\over n+2}\sum_{p=1}^{n-1}b_{n-p}\,c_p,\quad
n>0\label{cee}
\ee
Similarly $\Xi_4$ can be obtained from the term in the \Sc equation that is quartic in $\BA^\rho$

\bq&&\lim_{\lambda\rightarrow\infty}R'_\lambda\,\Big(
{1\over g_0^2}\BA^\rho\cdot(\BA^\rho\cdot(\BA^\rho\cdot(\BA^\rho\cdot \dot\Xi_4)))
\nn&&\quad\quad
 -8\BA^\rho\cdot(\BA^\rho\cdot((\BA^\rho\cdot\Xi_4)))\cdot(\BA^\rho\cdot\Xi_2)     -{9\over 2}\BA^\rho\cdot((\BA^\rho\cdot\Xi_3))\cdot\BA^\rho\cdot((\BA^\rho\cdot\Xi_3))\nn
&&\quad\quad      +{1\over 2g_0^4}\BA^\rho\cdot(\BA^\rho\cdot(\BA^\rho\cdot(\BA^\rho\cdot V_4)))
\Big )=0
\label{sf4}
\eq
Taking $\Xi_4$ to have the form

\be
\Xi_4={\tau\over g_0^2}\sum_{n=0}^\infty f_n (\tau^2 V_2\cdot)^n V_4
+{\tau\over g_0^2}\sum_{p,q,r} h_{p,q,r} \tau^{2(p+q+r+1)} (V_2\cdot)^p
\left(\left((V_2\cdot)^qV_3\right)\cdot\left((V_2\cdot)^rV_3\right)\right) 
\label{xi4}
\ee
leads eventually to the following formulae for the coefficents:

\be
f_0=-{1\over 10},\quad f_n={8\over 2n+5}\sum_{p=1}^{n-1}b_{n-p}\,f_p,\quad
n>0\label{eff}
\ee
and
\be
h_{0,q,r}={9\over 4(q+r)+14}c_qc_r,\quad h_{n,q,r}={8\over 2(n+q+r)+7}\sum_{p=0}^{n-1}b_{n-p}\,h_{p,q,r},\quad
n>0\label{eff2}
\ee

\section{Heat-Kernel Results}

In \cite{PYM} it was shown that

\be{\partial\over\partial s}\,Tr\,{ K}=
{\partial\over\partial s}\left(Tr\,exp\,\left({s(\DC\cdot\DC {\bf 1}+
2{\bf F})\cdot}\right)- Tr\,exp\,\left({s\DC\cdot\DC}\right)\right).
\ee
where for background fields that are slowly varying on the scale of $s$

\be
exp\,\left({s(\DC\cdot\DC {\bf 1}+
2{\bf F})\cdot}\right)^{CD}\delta (\xx -\yy)\,{\bf 1}
={e^{-(\xx -\yy )^2/(4s)}\over {\sqrt{4\pi s}}^3}\sum_{n=0}^\infty
s^n{\bf b}_n^{BC}(\xx ,\yy ) 
\ee
and 
\be
exp\,
\left( {s\DC\cdot\DC}\right)^{CD}\delta (\xx -\yy
)={e^{-(\xx -\yy )^2/(4s)}\over {\sqrt{4\pi s}}^3}\sum_{n=0}^\infty
s^n{a}_n^{BC}(\xx ,\yy ) 
\ee
The first few coefficents are, at coincident argument,

\be
a_1(\xx ,\xx )=0,\quad a_2(\xx ,\xx )={1\over 12}F_{ij} F_{ij}
\ee
and
\be
{\bf b}_1(\xx ,\xx )=2{\bf F},\quad
{\bf b}_2 (\xx ,\xx )=2{\bf F}\cdot{\bf F}+{1\over 12}{\bf 1}\,
F_{ij}F_{ij}+{1\over 3}\DC\cdot\DC {\bf F}
\ee
Resulting in 
\be
{\partial\over\partial s}\,Tr\,{ K}={11N\over 3}{1\over (4\pi)^{3/2}}
{1\over \sqrt s}{\cal B}_0+O(\sqrt s).
\ee
which gives

\bq
s{\partial \over\partial s}\Delta_s\,\ba\cdot(\ba\cdot\Xi)\Big|_{{\cal B}_0}
&=&{22N\over 3}{1\over (4\pi)^{3/2}}
\sum_{n=0}^\infty b_n \,{s\over\tau}\left(-\tau^2 {\partial\over\partial s}\right)^n
{1\over \sqrt {s}}\nn
&=&{22N\over 3}{1\over (4\pi)^{3/2}}
\sum_{n=0}^\infty b_n {\Gamma (n+1/2)\over \Gamma(1/2)}\left({\tau^2\over s}\right)^{n-1/2}
\eq


\begin{thebibliography}{88}
\bibitem{SYMANZIK} Symanzik, K.  Nucl. Phys. B190[FS3] (1983) 1

\bibitem{REV5}Symanzik, K. {\it Schrodinger Representation in
Renormalizable Quantum Field Theory},
Les Houches 1982, Proceedings, Recent Advances In Field Theory and
Statistical Mechanics

\bibitem{REV4}Luscher, M. {\it Schrodinger Representation in Quantum
Field Theory}, Nucl. Phys. B254 ( 1985) 52-57.

\bibitem{Paul1} Mansfield, P., Phys. Lett. B358 (1995) 287

\bibitem{Green}Greensite, J.,  Nucl.Phys. B158 (1979) 469,
Nucl.Phys. B166 (1980) 113

\bibitem{SINT} Sint, Sommer, {\it One loop renormalisation of QCD \Sc functional},
Nucl. Phys. B451 (1995) 416.

\bibitem{LUSCHER} L\"uscher, M. , Narayanan, R. , Weisz, P. and Wolff,U.
, Nucl. Phys. B384 (1992) 168


\bibitem{PMJ}Mansfield, P., Pachos, J., Sampaio, M., {\it Short Distance Properties
from Large Distance Behaviour}\,  Int. J. Mod. Phys. A., {\it to appear.}

\bibitem{PM}Mansfield, P., {\it Reconstructing the Vacuum
Functional of Yang-Mills from its Large Distance Behaviour},
Phys. Lett. B365 (1996) 207.

\bibitem{forth}Mansfield, P., {\it Surface Critical Scaling and the \Sc Equation}, in preparation.


\bibitem{PYM} Mansfield, P., Nucl. Phys. B418 (1994) 113


\bibitem{Zarembo} Zarembo, K., hep-th/9803237,9804276



\bibitem{Hat} Hatfield, B. {\it Quantum Field Theory of Particle and
Strings}, Addison Wesley, 1992, ISBN 0-201-11-11982X

\bibitem{CHAN} Chan, H.S., Nucl. Phys. B278(1986)721


\bibitem{Jac} Jackiw, R. {\it Analysis on Infinite Dimensional Manifolds:
Schrodinger
Representation for Quantized Fields}, Brazil Summer School 1989:78-143

\bibitem{Hugh} McAvity, D.M. and Osborn, H. Nucl. Phys. B394 (1993) 728

\bibitem{Ji} Oh, P., Pachos, J., hep-th/9807106.

\bibitem{REV2}Yee, J. H., {\it Schrodinger Picture Representation of
Quantum Field Theory}, Mt. Sorak Symposium 1991:210-271.

\bibitem{REV3}Floreanini, R.,{\it Applications of Schrodinger Picture in
Quantum Field Theory}, Luc Vinet (Montreal U.),  MIT-CTP-1517, Sept. 1987.


\bibitem{Kief1} Kiefer, C., Phys. Rev. D 45 (1992) 2044

\bibitem{Kief2} Kiefer, C., Wipf, A., Ann. Phys. 236 (1994)241


\bibitem{Feyn}Feynman, R. P., Nucl. Phys. B188(1981) 479


\bibitem{MMaeda}Kawamura, M., Maeda, K., Sakamoto, M., KOBE-TH-96-02
hep-th/9607176


\bibitem{IS1} Dunne, G.V., Jackiw, R., Trugenberger, C.A., {\it Chern-Simons
Theory in the Schrodinger Representation}, Ann.Phys.194:197,1989.

\bibitem{IS2} Ramallo, A.V.,{\it Two Dimensional Chiral Gauge Theories in
the Schrodinger
Representation}, Int.J.Mod.Phys.A5:153,1990.


\bibitem{else}  Heck, K. {\it  Some Considerattions on the Problem of
Renormalization of Quantum
Field Theory in the Schrodinger Representation} In German, Heidelberg
Univ. - HD-THEP 82-04 .




\bibitem{Horiguchi} Horiguchi, T.,KIFR-94-01,
KIFR-94-03, KIFR-95-02, KIFR-96-01, KIFR-96-03.

\bibitem{Horiguchi2} Horiguchi, T., Nuovo. Cim. 111B (1996) 49,85,293.


\bibitem{HMS} Horiguchi, T., Maeda, K., Sakamoto, M., Phys.Lett. B344 (1994) 105.

\bibitem{JK} Kowalski-Glikman, J., Meissner, K.A., Phys.Lett. B376 (1996) 48.

\bibitem{JK2} Kowalski-Glikman, J. gr-qc/9511014.

\bibitem{JK3} Blaut, A.,Kowalski-Glikman, J. gr-qc/9607004

\bibitem{KARA} Karabali, K., Kim C., Nair V.P., {\it On the vacuum wavefunction and string tension of Yang-Mills theories in (2+1) dimensions}, CCNY-HEP 98/3
RU-98-6-B SNUTP 98-034.

\bibitem{Maeda} Maeda, K., Sakamoto, M., Phys.Rev. D54 (1996) 1500.


\bibitem{witten}Witten, E.,{\it Anti de Sitter Space and Holography},
hep-th/9802150.









\end{thebibliography}
\end{document}